\documentclass[preprint,12pt]{elsarticle}

\usepackage{graphicx}
\usepackage{epstopdf}
\usepackage{mathptmx}      
\usepackage{amsmath}
\usepackage{float}
\usepackage{enumerate}
\usepackage{url}
\usepackage{latexsym}
\usepackage{mathtools}
\usepackage{bm}
\usepackage{esvect}
\usepackage{appendix}
\usepackage[svgnames,x11names]{xcolor}
\usepackage{color}
\usepackage{multirow}
\usepackage{threeparttable}
\usepackage{booktabs}
\usepackage{verbatim}
\usepackage{subfigure}
\usepackage{makecell}
\usepackage{fixmath}
\usepackage{rotating}
\usepackage{fix-cm}
\usepackage{mathrsfs}
\usepackage{amssymb}
\usepackage{etoolbox}
\usepackage{amsthm}
\usepackage[bookmarks=true]{hyperref}
\usepackage{lineno}
\usepackage{nomencl,etoolbox,ragged2e,siunitx} 

\newcommand\Reviewfirst[1]{\textcolor{Black}{#1}}
\newcommand\Reviewsecond[1]{\textcolor{Black}{#1}}
\newcommand\Fischer[1]{\textcolor{Black}{#1}}
\emergencystretch=\maxdimen
\newcommand{\DimensUnits}[1]{\hfill\makebox[3em]{#1\hfill}\ignorespaces}
\newcommand{\insertnomheaders}{\item[\bfseries Symbol]%
\textbf{Description}\DimensUnits{\textbf{Units}}}
\renewcommand\nomgroup[1]{\insertnomheaders}

\journal{Experimental Thermal and Fluid Science}

\begin{document}

\begin{frontmatter}



\title{{Measurement error of tracer-based velocimetry in single-phase turbulent flows with inhomogeneous refractive indices}}

\address[mymainaddress]{Center of Applied Space Technology and Microgravity (ZARM), University of Bremen, 28359 Bremen, Germany}
\address[mysecondaryaddress]{The State Key Laboratory of Nonlinear Mechanics, Institute of Mechanics, Chinese Academy of Sciences, 100190 Beijing, China}
\address[mythirdaryaddress]{University of Bremen, Bremen Institute for Metrology, Automation and Quality
Science (BIMAQ), 28359 Bremen, Germany}
\address[myquaternaryaddress]{MAPEX Center for Materials and Processes, University of Bremen, 28359 Bremen, Germany}

\author[mymainaddress]{Huixin Li} 

\author[mythirdaryaddress,myquaternaryaddress]{Andreas Fischer}

\author[mymainaddress,myquaternaryaddress]{Marc Avila}

\author[mymainaddress,mysecondaryaddress]{Duo Xu\corref{cor1}}
\cortext[cor1]{Corresponding author}
\ead{duo.xu@imech.ac.cn}

\begin{abstract}
Inhomogeneous refractive index fields lead to errors in optical flow velocity measurements. Former respective studies are mostly in quasi two-dimensional flows, and attribute the measurement errors to spatial gradients in the refractive index field, while less attention has been paid to flows with three-dimensional refractive index fields which usually change in space and in time. In this study, ray tracing simulations were carried out in a three-dimensional flow, which is from a direct numerical simulation of single-phase turbulent mixing of two fluids. 
Given the data of the numerical simulation as reference, the ray tracing simulation is used to quantify the measurement errors of the flow velocity and flow acceleration for tracer-based velocimetry, i.e. particle tracking velocimetry in this study. The errors of both flow velocity and flow acceleration are attributed to the spatial and the spatio-temporal gradients of the refractive indices, respectively, which are closely inherited from flow characteristics. \Reviewfirst{\Fischer{While the dominant type of error depends on the studied flow, the main measurement error for the considered turbulent mixing flow is caused by the random error.}} When the maximum spatial difference of the refractive indices is about $10^{-6}$, the relative random measurement error is about 1 \% in velocity and about $200$ \% in acceleration, respectively. \Reviewfirst{When the maximum index difference is about $10^{-2}$ \Fischer{(water)}, the relative random measurement errors of velocity and acceleration are $2000$ \% and $10^5$ \%, respectively, for the flow considered in this study.}
\end{abstract}

\begin{graphicalabstract}
\includegraphics[width=1\textwidth]{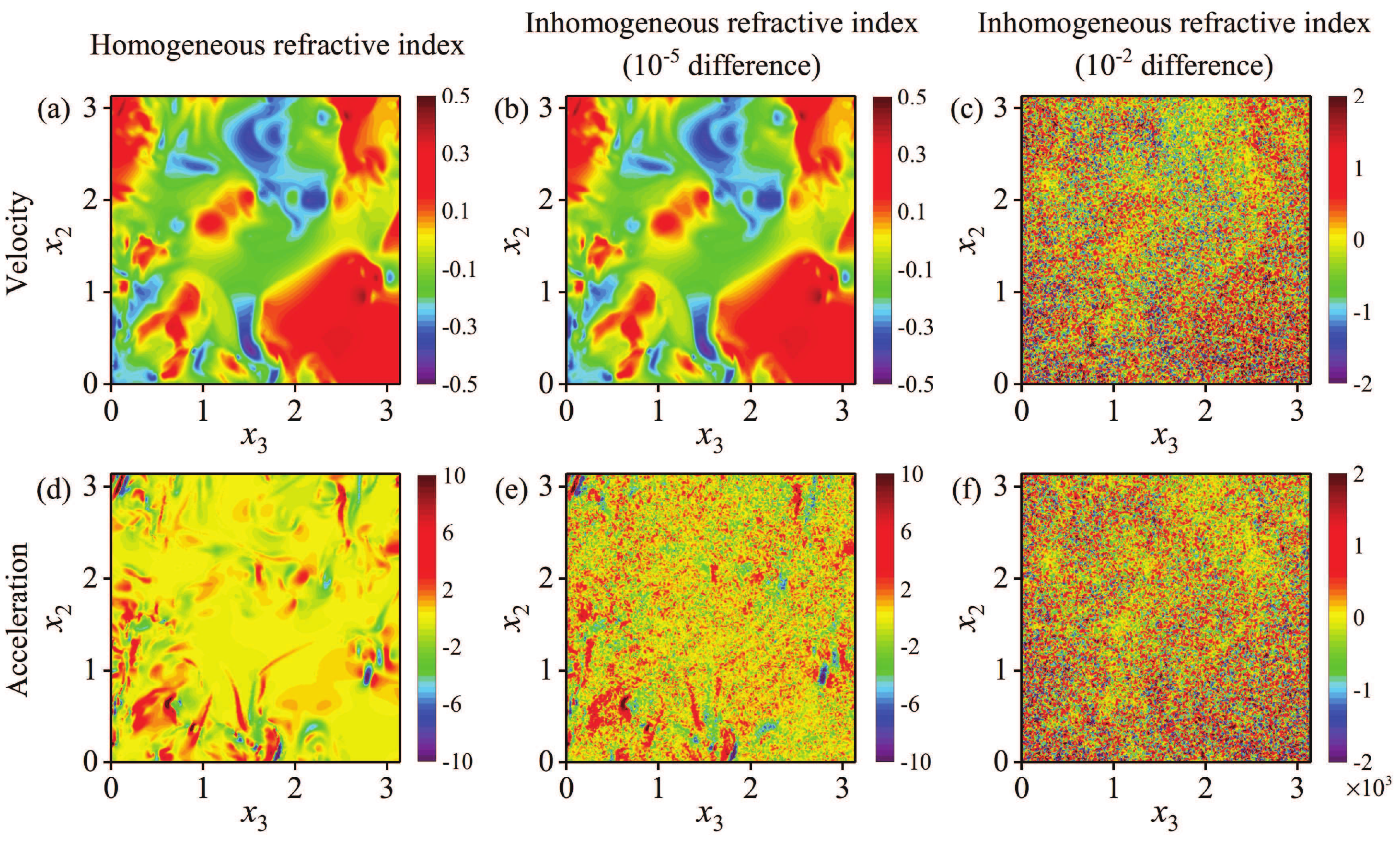}
\end{graphicalabstract}

\begin{highlights}
\item {\Reviewfirst{Ray tracing simulations were carried out to quantify PTV measurement errors.}}
\item {\Reviewfirst{The refractive index field was obtained from the density data of a \Fischer{DNS} dataset.}} 
\item {\Reviewfirst{The maximum spatial difference of refractive indices ranges from $10^{-6}$ to $10^{-2}$.}} 
\item{\Reviewfirst{The random error is the dominant for the velocity and acceleration for the \Fischer{studied flows}.}} 
\item {\Reviewfirst{The velocity error is associated with \Fischer{spatio-temporal} gradients of refractive indices.}}
\end{highlights}


\end{frontmatter}


\makenomenclature
\providetoggle{nomsort}
\settoggle{nomsort}{true} 

\makeatletter
\iftoggle{nomsort}{%
	\let\old@@@nomenclature=\@@@nomenclature        
	\newcounter{@nomcount} \setcounter{@nomcount}{0}%
	\renewcommand\the@nomcount{\two@digits{\value{@nomcount}}}
	\def\@@@nomenclature[#1]#2#3{
		\addtocounter{@nomcount}{1}%
		\def\@tempa{#2}\def\@tempb{#3}%
		\protected@write\@nomenclaturefile{}%
		{\string\nomenclatureentry{\the@nomcount\nom@verb\@tempa @[{\nom@verb\@tempa}]%
				\begingroup\nom@verb\@tempb\protect\nomeqref{\theequation}%
				|nompageref}{\thepage}}%
		\endgroup
		\@esphack}%
}{}
\makeatother

\section{Introduction}\label{sec:intro}
{In experimental fluid mechanics, tracer-based optical measurement techniques bring in substantial insights into the physics of flows \citep[]{adrian1991piv, Maas1993ptv}. The typical examples of such measurement techniques are particle image velocimetry (PIV) and particle tracking velocimetry (PTV), which share the same working principle. To implement PIV/PTV measurements, the fluid is seeded with small tracers with negligible buoyancy and inertia. With a powerful illumination, the spatial positions of the illuminated tracers are recorded by the camera(s).} Assuming negligible slip, the flow velocity equals the tracer velocity and can be determined from the correlation of recorded image pairs \citep{raffel2018particle}. 
In flows with a homogeneous refractive index field, a flow velocity measurement error of about 1~\% is typically feasible \citep{raffel2018particle}, for a sufficient quality of the tracer images.
However, the image quality may not be reached in inhomogeneous refractive index fields with spatio-temporal variations, which occur for instance in flows with shock waves, combustion, thermal convection and fluid mixing. Here, a photon does not travel along a straight path, but follows a more complex trajectory (light ray), e.g., due to light refraction and diffraction. Consequently, the image of a tracer particle can be blurred and a position error of the tracer in the image can occur. The deterioration of the tracer image quality then results in an increased measurement error of the flow velocity.

In compressible air flows, the position error and the blur of the tracer image can be dramatic when the illuminated field is seen through shock waves or shear layers \citep{raffel1998investigation,elsinga2005evaluation}. In liquid flows, the optical distortion was, e.g., observed in fluorescence images of scalar mixing from two water streams with different temperature \citep{oljaca2009effects}. In turbulent flames, measurements with PIV and laser Rayleigh imaging technique are also affected by the inhomogeneous refractive index field produced by temperature differences \citep{stella2001application,kaiser2005use}. {In flows of air-water free surface, the position error of tracers due to light refraction at the free surface was recorded, and it was used to reconstruct the surface topographic structures \citep{moisy2009synthetic,gomit2013free}.} Analogously, laser beam deflection, the image distortion and the PIV measurement error are also reported for a hot jet, porous media flows (with the refractive index difference between the solid and the liquid),  and a thermal boundary layer of a melting paraffin wax, respectively  \citep{vanselow2018influence,patil2012optical,faden2019velocity}.

Many studies have focused on flows with steady inhomogeneous refractive index fields, e.g., the shock waves in aero-optical studies of a seeker on a hypersonic conical vehicle \citep{guo2016aero}. A shock wave has a discontinuity of the {refractive index} field produced by the density (temperature) fields, and is often approximately two-dimensional. \citet{raffel1998investigation} introduced a formula to estimate {the size enlargement of the imaged tracers} in order to quantify the position error of the imaged tracers seen across the shock wave. \citet{elsinga2005evaluation} studied aero-optical errors of PIV measurements in an approximately two-dimensional flow and found that the formula of \citet{raffel1998investigation} overestimates the {light ray} deflection. For the approximately two-dimensional shock wave attached to a conical-head vehicle, \citet{guo2016aero} simulated light rays over discrete grids following Snell's law. However, the numerical iteration of Snell's law over discrete grids possibly gives an insufficient accuracy, if grid sizes are not sufficiently small \citep{stam1996raytracing}. In contrast, solving the Fermat's ordinary differential equation, e.g., with a Runge-Kutta scheme, provides a good accuracy in {ray tracing} simulations \citep{kirmse2011application}. This method can render almost realistic PIV and background-oriented Schlieren images in a two-dimensional flow \citep{rajendran2019piv}. 

\begin{table}
	\centering
	\caption{The list of studies on tracer position error (or angle deflection) from inhomogeneous index field. }\label{tab:studies}
	\small
	\renewcommand{\arraystretch}{1}
	\begin{threeparttable}
		\begin{tabular}{p{2.15cm} p{2.35cm} p{1.25cm} p{1.75cm} p{1.75cm} p{2cm} }
			\toprule
			Reference & Flow & Medium & $\Delta n_\mathrm{max}$ & $K$ (m$^3$/kg) & Method  \\ \midrule
			\citet{vanselow2018influence} \tnote{1} & Hot jet flow & Air & $9.9 \times 10^{-5}$ & 2.3 $\times 10^{-4}$ & Experiment: laser beam  \\
			\citet{guo2016aero} \tnote{2} & Shock wave on conical vehicle & Air & $2.0 \times 10^{-5}$ & 2.3 $\times 10^{-4}$ & Simulation: Snell's law (DSMC) \\
			\citet{elsinga2005evaluation} \tnote{3} & Compressible shear layer & Air & $9.0 \times 10^{-5}$ & 2.3 $\times 10^{-4}$ & Theory  \\
			&  & Air & $6.9 \times 10^{-5}$ & 2.3 $\times 10^{-4}$ & Experiment: PIV \& BOS  \\
			& Prandtl-Meyer expansion fan & Air & $7.4 \times 10^{-5}$ & 2.3 $\times 10^{-4}$ & Theory  \\
			&  & Air & $7.1 \times 10^{-5}$ & 2.3 $\times 10{^-4}$ & Experiment: PIV \& BOS \\
			\citet{stella2001application} \tnote{4} & Premixed turbulent flames & Air-propane & $1.3 \times 10^{-3}$ & $3.0 \times 10^{-4}$ & Theory  \\
			&  & Air-propane & &  & Experiment: laser beam  \\
			\citet{raffel1998investigation} \tnote{5} & Shock wave & Air & $1.0 \times 10^{-4}$ & $2.3 \times 10^{-4}$ & Experiment: Snell's law \\
			\citet{kirmse2011application} \tnote{6} & Shock wave & Air & $6.3 \times 10^{-6}$ & $2.5 \times 10^{-4}$ & {Simulation: ray tracing (CFD).\quad Experiment: BOS}  \\
			\citet{oljaca2009effects} \tnote{7} & Plane shear layer & Water & $3.4 \times 10^{-4}$ & -- & Experiment: LIF \\
			\bottomrule
		\end{tabular}
		\begin{tablenotes}
			\footnotesize
			\item[1] Gradient of refractive indices along the direction perpendicular to the laser beam was converted and used here. 
			\item[2] Incidence angle was not explicitly defined.
			\item[3] Refractive index field was extracted from their figure~9 and 10.
			\item[4] Refraction at flame fronts was considered. The data from their table~5 were extracted.
			\item[5] The shock wave was assumed as the interface of two media with refractive index difference.
			\item[6] Maximum refractive index difference was extract from the legend of their figure~12.  
			\item[7] Dependence of deflection angels on spatial distribution of refractive index field was discussed but no data were presented. The temperature difference is up to $5 ^{\circ}$C.
		\end{tablenotes}
	\end{threeparttable}
\end{table}

Much less attention has been paid to flows with unsteady inhomogeneous refractive index fields. One example of such flows is the Rayleigh-B\'enard convection in a closed cell, which is heated at the bottom and cooled at the top \citep{bodenschatz2000RB,lohse2010smallRB}. The characteristic three-dimensional flow structures (e.g. plumes and large-scale circulation) advect in time \citep{bodenschatz2000RB} with velocity and temperature fluctuating strongly at small scales \citep{lohse2010smallRB}. In the meanwhile, correspondingly, the three-dimensional {refractive index} patterns deflect the light in space, and importantly the deflection also changes in time. The variation of the light deflection over time brings in another dimension of the  deterioration of the {tracer image quality}. 

The photon trajectories in these flows are determined by the spatio-temporal behaviour of the refractive index field inside the flow \citep{vanselow2018influence}. In a recent experimental study, \citet{Vanselow2019} quantified the standard PIV measurement error in a combustion flow. In their study, the tracer position error inside the flow was measured in a time-averaged manner. This averaged error of the tracer position was then combined with the time-averaged flow velocity from the PIV measurement inside the flame to determine the PIV measurement error. \citet{Vanselow2019} found that the \emph{time-averaged} relative velocity error over 500 PIV measurements is up to $4\ \%$. This systematic measurement error is larger than the typical relative error of about 1 \% for a single PIV measurement \citep{Westerweel1997,raffel2018particle}. More recently, they extended their study to the velocity measurement errors for stereoscopic PIV in the same experimental setup and found that the measurement error can be one order of magnitude larger than the standard PIV~\citep{vanselow2021stereoscopic}.
Moreover, the temporal evolution of the flow, producing changes to the {refractive index} field, influences the photon trajectories, even if the photon starts at the same position and along the same direction. The time-dependent error of the tracer position and the velocity error are coupled \citep{elsinga2005evaluation,Vanselow2019}, thus the position error and the actual velocity are simultaneously required for the quantification of the \emph{instantaneous} velocity error to evaluate the systematic and the random error of the measurements. However, this is difficult (or infeasible) in experiments.

As an alternative approach, the effect of the inhomogeneous refractive index field, which varies spatio-temporally on the measurement error, can be studied with ray tracing simulations in simulated flows \citep{kirmse2013investigation}, where the {photon trajectories inside the flows can be simulated on the basis of the known refractive index field in the scope of geometric optics (neglecting light diffraction)}. Moreover, {this approach enables studying} three-dimensional refractive index fields, which is particularly important for turbulent flows since being chaotic, three-dimensional and feature substantial scale interactions. However, a respective study of the measurement error of the flow velocity and the acceleration {for tracer-based measurement principles} is missing.

In contrast to PTV measurements, a velocity value measured with PIV corresponds to a {most probable} velocity of several tracers in an interrogation window, which is approximately regarded as a spatial averaging effect over the velocities of the tracers in the window \citep{xu2013}. In order to study the effect of the spatio-temporally varying refractive index field on the measured tracer motion only, the present article is focused on the PTV measurement error of the flow velocity and the flow acceleration, respectively, which are investigated in a simulated single-phase, three-dimensional turbulent flow. 
In Section~\ref{sec:methods}, we first describe in detail the measurement arrangement, and the velocity and acceleration error which are induced by the optical distortion in a refractive index field. Section~\ref{sec:setup} presents the implementation of respective numerical experiments in a varying refractive index field, including the methodology of generating synthetic images and simulation assumptions. The measurement errors of the tracer position, flow velocity and acceleration are elucidated in Section~\ref{sec:results}. Finally, Section~\ref{sec:conclusion and outlook} describes the conclusion and outlook. 

\section{Approach}\label{sec:methods}
\begin{figure*}
	\centering
	\includegraphics[width=0.95\textwidth]{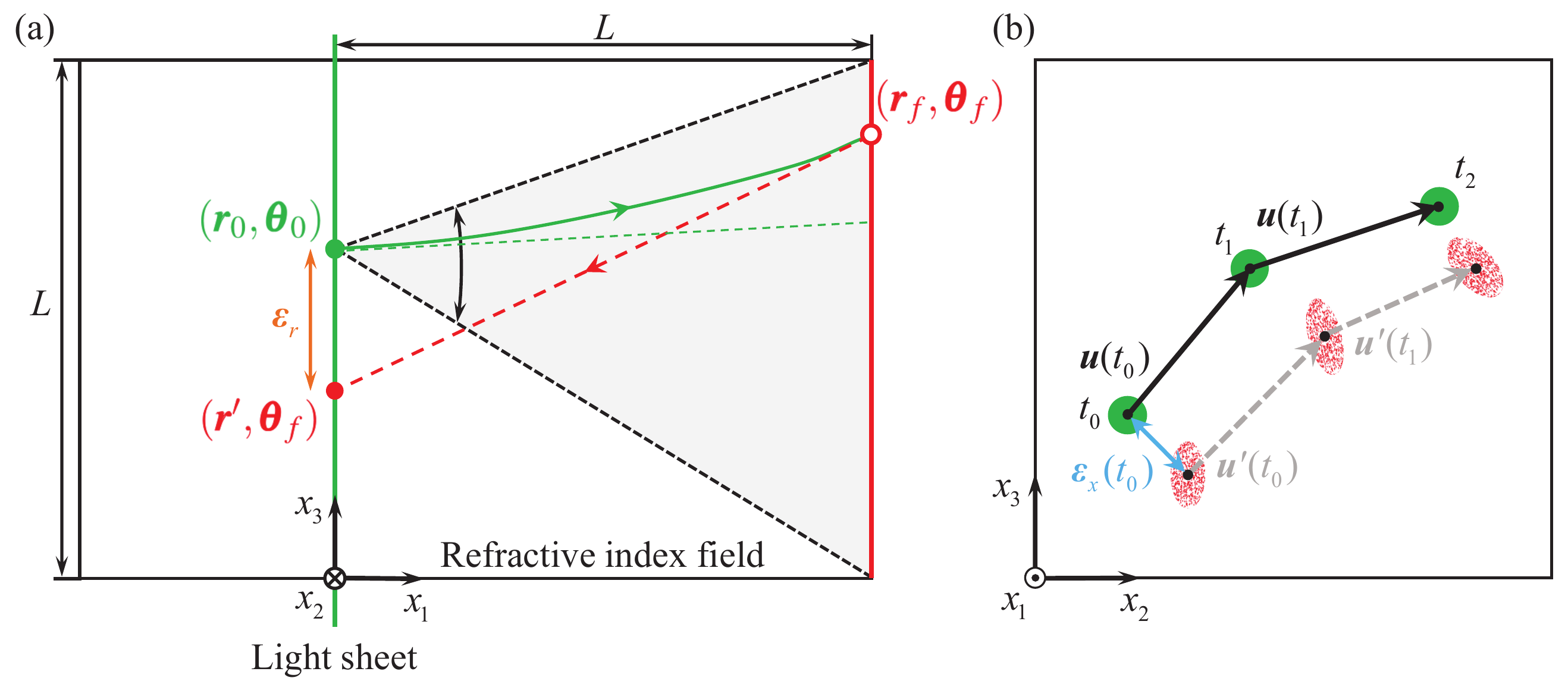}
	\caption{{(Color online) (a) Two-dimensional schematic of tracing a light ray in a three-dimensional inhomogeneous refractive index field. The gray area (enclosed by two black dashed lines) marks the range of $\pmb{\theta}_0$. The ray position shift is $\epsilon_r(=\pmb{r}_0-\pmb{r}')$ marked by an orange double-head arrow. (b) Illustration of tracer position and tracer velocity at three times ($t_0,t_1,t_2$) for the homogeneous and the inhomogeneous refractive index field cases. For the latter, the red dots at the position $\pmb{r}'$ in (a) form elliptical shaped tracers, while the green circles indicate the imaged tracers from the former case. Black dots mark the respective tracer center. Black and gray arrows indicate the respective velocity, and the blue double-head arrow marks the position error at $t_0$.}}\label{fig:sketch}
\end{figure*}

\subsection{Tracing light rays}\label{sec:distortion}
Tracing the light ray in a flow from the illuminated tracer is illustrated in figure~\ref{fig:sketch}(a). The flow domain is three dimensional in size of $L \times L \times L$. In the two-dimensional sketch, the $x_2-x_3$ plane at $x_1=0$ is illuminated with a light sheet (the green line). A illuminated tracer scatters a light ray at the position $\pmb{r}_0$ (the green dot) towards the plane $x_1=L$, and the initial direction of this ray (the green dashed line) is defined by a vector $\pmb\theta_0$ of angles in reference to $x_1$. In the scope of geometric optics, the propagation trajectory of a photon can be described by Fermat' s equation as
\begin{equation}\label{eq:rayeq}
\frac{\mathrm d}{\mathrm{d} s}[n(\pmb{r})\frac{\mathrm{d} \pmb{r}}{\mathrm{d} s}]=\nabla n(\pmb{r}),
\end{equation}
where $\pmb{r}$ is the ray curve, $n(\pmb{r})$ is the refractive index field, and $\mathrm{d}s$ is an infinitesimal increment of arc length along the trajectory, as well as $\nabla$ is a spatial gradient operator. {$n(\pmb{r})$ is obtainable from a direct numerical simulation (DNS) of a flow, so that a ray curve in the flow can be computed numerically in Fermat's equation when the initial conditions of the ray are given.}
In practice, light rays leaving the domain border of the flow field are collected by a group of optical lenses to a camera screen. The lenses may produce aberrations \citep{guenther2015modern}, which give difficulties to isolate the effect of the refractive index field on quantifying the measurement error. As shown in figure~\ref{fig:sketch}(a), in this study, a ray (the green line with an arrow) reaching the final position (the hollow red circle) in the final plane (red solid line) was projected back to the light sheet plane (the green vertical line) along a straight line (the red dashed line with an arrow) with the angle $\pmb\theta_f$. This configuration ensures that the studied ray deflection and the tracer position error are only associated with the refractive index field \citep{elsinga2005evaluation}. The projected ray reaches its destination $\pmb{r}'$ (the red dot in figure~\ref{fig:sketch}a) at $x_1=0$ (the light sheet), and a sufficient number of rays consequently forms an imaged tracer (see figure~\ref{fig:sketch}b). 

\subsection{Measurement arrangement}
Given the working principle of PTV techniques, the measurement precision is closely associated with the quality of tracer images, which is linked with the position of the imaged tracers that are formed by light rays. For this, the direction of the light ray, and then the position of the imaged tracers, as well as the flow velocity and the acceleration are evaluated.

The direction of the light ray, characterized by $\pmb{\theta}_0$ and $\pmb{\theta}_f$, is directly obtained from the ray tracing simulation. When a large number of the light rays from a tracer reaches $x_1=0$ (the imaging plane, the same as the light plane), an image of the tracer is rendered (see details in Section~\ref{subsec:measurement}). The measured tracer position $\pmb{x}'$ is the tracer center obtained by the centroid method for the image of a tracer,
\begin{eqnarray}
\pmb{x}'={\int_0^{L}\int_0^{L} \pmb{g}'(x_2,x_3)\cdot w(x_2,x_3) \; \mathrm{d} x_2 \mathrm{d} x_3}\left/\left(\int_0^{L}\int_0^{L} w(x_2,x_3) \; \mathrm{d} x_2 \mathrm{d} x_3\right)\right.,
\end{eqnarray}
where $\pmb{g}'$ are pixel coordinates in the image (see Section~\ref{subsec:measurement}), and $w$ is the weighting factor which is the pixel grayscale in this study. 

{The tracer velocity is obtained from the change of the tracer position $\pmb{x}'(t)$ in a sufficiently short time interval $\Delta t$:   }
\begin{equation}\label{eq:velocity}
\pmb{u}'(t)=\mathrm{d} \pmb{x}'(t)/\mathrm{d}t\approx[\pmb{x}'(t+\Delta t) - \pmb{x}'(t)]/\Delta t,
\end{equation}
see figure~\ref{fig:sketch}(b). In this study, the Lagrangian tracking of a tracer is considered, so that three positions of an individual tracer, rendered at three time instants in sequence, are used to evaluate the acceleration
\begin{eqnarray}\label{eq:acceleration}
\pmb{a}'(t) &=& \mathrm{d}\pmb{u}'(t)/\mathrm{d}t \\\nonumber &\approx& [\pmb{u}'(t+\Delta t) - \pmb{u}'(t-\Delta t)]/\Delta t \\\nonumber &\approx& [\pmb{x}'(t+\Delta t) - 2\pmb{x}'(t) + \pmb{x}'(t-\Delta t)]/\Delta t^2.	
\end{eqnarray}	

\subsection{Determination of the measurement error}
Regarding the measurement quantities above, we investigate the deflection of the light, the position error of the tracers and the velocity measurement error, as well as the acceleration measurement error.

The deflection of the light ray is represented by the direction difference $\Delta\pmb \theta={\pmb \theta}_f - {\pmb \theta}_0$, a quantity commonly used to evaluate aero-optical effects \citep{guo2016aero,jumper2017physics}. When the refractive index is homogeneous, $\Delta \pmb \theta=0$.
The position error between the measured position of the tracer and the true position (free of the inhomogeneous refractive index effect) is given by
\begin{equation}\label{eq:posi_error}
\pmb{\epsilon}_{x}(t)=\pmb{x}'(t) - \pmb{x}(t),
\end{equation}
where $\pmb{x}'$ is the tracer position measured in the inhomogeneous refractive index field and $\pmb{x}(=\iint \pmb{g}\cdot w \; \mathrm{d} x_2 \mathrm{d} x_3/(\iint w \; \mathrm{d} x_2 \mathrm{d} x_3))$ is the tracer position obtained based on the pixel coordinates ($\pmb{g}$) in the tracer image rendered from the homogeneous refractive index field. 

{The velocity error is quantified by the difference between the velocity measured in the inhomogeneous refractive index field $\pmb{u}'$ and the one in the homogeneous refractive index field $\pmb{u}$, \begin{equation}\label{eq:velocity_err}
\pmb{\epsilon}_u(t) = \pmb{u}'(t) - \pmb{u}(t),
\end{equation}
where $\pmb{u}$, the ground truth, is obtained through a cubic interpolation over the data of the DNS at the nearest eight neighbor grid points. Similarly, the acceleration measurement error is obtained by, 
\begin{equation}\label{eq:acce_err}
\pmb{\epsilon}_{a}(t) = \pmb{a}'(t) - \pmb{a}(t),
\end{equation}
where $\pmb{a}'$ and $\pmb{a}$ are the measured acceleration and the true value, respectively. 
The acceleration $\pmb{a}'$ is calculated with equation~\eqref{eq:acceleration} using the measured velocity $\pmb{u}'$. The true acceleration is not available in DNS, and it is computed also according to equation~\eqref{eq:acceleration} using the DNS velocity $\pmb{u}$ instead.

\section{Setup of the numerical experiments}\label{sec:setup}
In order to simulate the light rays in the refractive index field, the velocity and the density data of a DNS are used to carry out the numerical experiments.

\subsection{Simulated measurement object}\label{sec:dns}
The DNS data used for our ray tracing numerical experiments in this study are from the simulations performed by \citet{livescu2007buoyancy}. They implemented a DNS of homogeneous, buoyancy driven turbulence in a cube with periodic boundary conditions ($1024$ Fourier modes were used along each dimension), and zero-mean velocity and constant mean pressure gradient are imposed. They solved the incompressible Navier-Stokes equations of miscible two-fluid in single phase using a pseudo-spectral method and the Adams--Bashforth--Moulton scheme coupled with a pressure projection method. The equations were made dimensionless with density $\rho_\mathrm{fluid}=(\rho_1+\rho_2)/2$, velocity $U_0$ and reference length $L_0$ (leading to the cube edge of $2\pi$), where $\rho_1$ and $\rho_2$ correspond to the density of light and heavy fluids, respectively. $U_0$ and $L_0$ are not specified explicitly in \citep{livescu2007buoyancy}. The simulation was initialized with randomly distributed blobs of fluids, {then turbulence was} produced by the different buoyancy of the two fluids. The Reynolds number was $Re=\rho_\mathrm{fluid} L_0U_0/\mu_\mathrm{fluid}=12500$, where the dynamic viscosity $\mu_\mathrm{fluid}$ was the same for both fluids. The Schmidt number was $Sc=\mu_\mathrm{fluid}/(\rho_\mathrm{fluid}D_0)=1$ with the diffusion coefficient $D_0$. The density ratio of two fluids was $\rho_2/\rho_1=1.105$. Their dataset is available through the Johns Hopkins University Turbulence Database \citep{JHTDB}. The dimensionless velocity $\pmb u$, density $\rho$ and spatial gradients of density $\nabla \rho$ were downloaded for a grid of $512^3$ points ($1/8$ of the full domain) in a dimensionless time interval $0.005~(L_0/U_0)$ around the time $11.400$ ({{see figure~\ref{fig:dns}b for a snapshot of the refractive index field}}), where the flow turbulent kinetic energy reaches the maximum. 

{The inhomogeneous refractive index field $n$ is obtained from the dimensionless density $\rho$ via the Gladstone-Dale equation} \citep{patil2012optical},
\begin{eqnarray}\label{eq:GD}
n=K_{\mathrm{fluid}}\,  \rho_{\mathrm{fluid}} \rho + 1, \quad \mathrm{and\;\; thus,} \quad 
\nabla n = K_{\mathrm{fluid}} \, \rho_{\mathrm{fluid}}\, \nabla\rho, 
\end{eqnarray}
where $K_\mathrm{fluid}$ is the Gladstone-Dale constant (depending on the fluid), and $\rho_{\mathrm{fluid}}$ is the (dimensional) fluid density. 
In our study, $K_{\mathrm{fluid}} \rho_{\mathrm{fluid}}$ is changed from $3.34 \times 10^{-5}$ to $3.34 \times 10^{-1}$ to represent five kinds of fluid, as shown in table~\ref{tab:cases}, where C1 and C4 correspond to air ($K_\mathrm{air}=2.3 \times 10^{-4}~\mathrm{m^3/kg}$, $\rho_\mathrm{air} = 1.225~\mathrm{kg/m^3}$) and water ($K_\mathrm{water}=3.34 \times 10^{-4}~\mathrm{m^3/kg}$, $\rho_{\mathrm{water}}=10^3~\mathrm{kg/m^3}$), respectively.

\begin{table}
\centering
\caption{Parameters in the simulation cases.}\label{tab:cases}
	\resizebox{\textwidth}{10mm}{
	\begin{threeparttable}
	\begin{tabular}{llllll}\toprule
Case                     & C0       & C1 (air flow)               & C2                           & C3                     & C4 (water flow)                     \\ \midrule
$K_{\mathrm{fluid}} \rho_{\mathrm{fluid}}$ & $3.34 \times 10^{-5}$ & $2.82 \times 10^{-4}$ & $3.34 \times 10^{-3}$  & $3.34 \times 10^{-2}$ & $3.34 \times 10^{-1}$               \\
$\Delta n_{\mathrm{max}}$      &  $2.85 \times 10^{-6}$        & $2.40 \times 10^{-5}$ & $2.85 \times 10^{-4}$  & $2.85 \times 10^{-3}$ & $2.85 \times 10^{-2}$ \\
\bottomrule
\end{tabular}
	\end{threeparttable}}
\end{table} 

\begin{figure}
	\centering
	\includegraphics[width=1\textwidth]{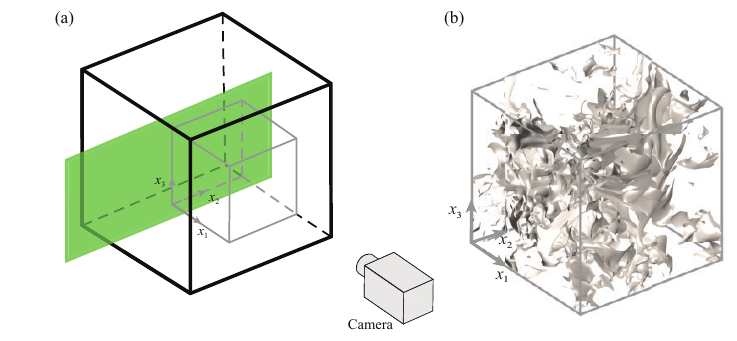}
	\caption{(Color online) (a) The illustration of the numerical experiment domain (gray box, $L \times L \times L$), part of the DNS domain (black box), where the green layer denotes a light sheet. The camera views the light sheet through the gray box. (b) An example of the isosurface of the refractive index field.}\label{fig:dns}
\end{figure}

\subsection{Simulation of the flow measurement}\label{subsec:measurement}
In this study, the numerical experiment was carried out in a cubic domain, the gray box in figure~\ref{fig:dns}(a).
The flow domain has a length $L$ along each of the three dimensions. A Cartesian coordinate $(x_1,x_2,x_3)$ is set with $x_1\in[0,L]$, $x_2\in[0,L]$ and $x_3\in[0,L]$.
{A two-dimensional measurements (for $x_2$ and $x_3$ components) are simulated at the plane ($x_1=0$, $x_2$, $x_3$).
The plane $x_1=0$ is illuminated with a light sheet {with a light} wavelength of $\lambda=532$~nm.} {In the light sheet, $10^5$ tracers in size of $10~\mu$m are distributed homogeneously and randomly, and they are assumed to move within the plane of the light sheet $x_1=0$ only.}
For each individual tracer, $\pmb{r}_0$ pointing towards $100\times100$ elements with equidistant grids at plane $x_1=L$ defines the initial direction $\pmb{\theta}_0$ of $10^4$ rays. 
Evaluating $10^4$ rays for each of the $10^5$ tracers is enough for the simulation of the tracer-based velocimetry \citep{rajendran2019piv}.
The simulations were carried out in a NVIDIA TITAN V GPU, which has $5120$ cores and the memory of $11.26$ GB. The simulation of $10^9$ rays takes approximately $12$ hours.

An in-house \textsc{Matlab} script was developed following the Runge-Kutta algorithm of \citet{sharma1982tracing} to trace each light ray in the refractive index field. In matrix form, the equation~\eqref{eq:rayeq} reads
\begin{equation}\label{eq:Numrayeq}
	\frac{\mathrm{d}^2 \pmb{R}}{\mathrm{d}\pmb{T}^2} = \pmb{D},
\end{equation}
where $\pmb{R}=(x_1, x_2, x_3)$ and $\pmb{T} = n (\mathrm{cos}\theta_1, \mathrm{cos}\theta_2, \mathrm{cos}\theta_3 )$ represent the position and the direction (with angle $\theta_i$ of a ray segment in reference to $x_i$), respectively, and $\pmb{D} = n (\partial n/ \partial x_1, \partial n/\partial x_2, \partial n/\partial x_3)$. The equation~\eqref{eq:Numrayeq} is solved iteratively in sequence as follows,
\begin{eqnarray}\label{RK}
	{[1]} &\quad \pmb{A}&=\quad \Delta \zeta \cdot \pmb{D}(\pmb{R}_j), \\ 
	{[2]} &\quad \pmb{B}&=\quad \Delta \zeta \cdot \pmb{D}(\pmb{R}_j+{\Delta \zeta}\;\pmb{T}_j/2+\Delta \zeta\; \pmb{A})/8, \nonumber \\
	{[3]} &\quad\pmb{C}&=\quad \Delta \zeta \cdot \pmb{D}(\pmb{R}_j+\Delta \zeta\;\pmb{T}_j+ \Delta \zeta\;\pmb{B}/2), \nonumber\\
	{[4]}&\quad\pmb{R}_{j+1}&=\quad \pmb{R}_j + \Delta \zeta \cdot [\pmb{T}_j+(\pmb{A}+2\;\pmb{B})/6], \nonumber\\
	{[5]}&\quad\pmb{T}_{j+1}&=\quad \pmb{T}_j + (\pmb{A}+4\;\pmb{B}+\pmb{C})/6, \nonumber
\end{eqnarray} 
where $\Delta \zeta$ is the simulation step size.
For each light ray, equation~\eqref{RK} is numerically iterated from $\pmb{R} =\pmb{r}_0$ and $\pmb{T}(n, \pmb\theta_0)$ at $x_1=0$ to the final plane $x_1=L$. In the iterations, $n(x_i)$ and $\nabla n(x_i)$ at the grids are available, elsewhere linear interpolation using neighboring eight vertices of a cubic volume was used to compute the sub-grid $n$ and $\nabla n$. {The simulation step size was tested and $\Delta \zeta=10^{-4}$ gives converged results.
This code was verified by simulations of two standard cases where their analytical solutions are available. The verification simulations and convergence tests are detailed in \ref{app:verification}.}

To render tracer images, ray physical coordinates are projected to image coordinates. The intensity of a light ray, given by the initial direction $\pmb{\theta}_0$ in the Mie scattering, is assumed to be unchanged along its path \citep{born_wolf1999}. The grayscale of each pixel is obtained by summing up the intensity of the rays reaching each pixel. The resulting `image' is then rescaled to a 10-bit dynamic range. A Gaussian filtering operation was performed to produce a tracer image to mimic the diffraction effect of the aperture \citep{rajendran2019piv}. A high image resolution of $8192^2~\mathrm{pixel}^2$ is used in this study \Reviewsecond{for resolving the small position change of the tracers images as a result of the inhomogeneous refractive index field. The size of tracer image in the non-distorted case (homogeneous refractive index field) is about 10~pixels (which corresponds to about $2-3$~pixels in a $2000^2$~pixels$^2$ imaging system in practice, which is a typical measurement condition). 
In a practical imaging system, particle images smaller than one pixel may give larger random error and lead to the known 'peak locking' issue (see Chapter 6 in \citet{raffel2018particle}), while too large tracer images may result in more tracer-tracer overlapping in the images.}
The truncation of the infinite number of rays in reality to a finite number for a tracer in a simulation produces an artificial effect to the image of a tracer, namely the scattered background noise resulted from the scattered ray destination positions in the image. Hence, a grayscale threshold is used to isolate a tracer from the background noise. For the air flow, the grayscale threshold is $100$ for the $10$-bit images, while for the water flow this threshold is $500$ due to large scatter in high-resolution images.
The examples of rendered tracer images are shown in figure~\ref{fig:render}, where the change of the tracer area is small in the air flow (C1), while in the water flow (C4) the tracers blur with large area changes, as visualized in figure~\ref{fig:render}(c). 

\begin{figure*}
	\centering
	\includegraphics[width=1.0\textwidth, trim=0cm 0cm 0cm 0cm, clip]{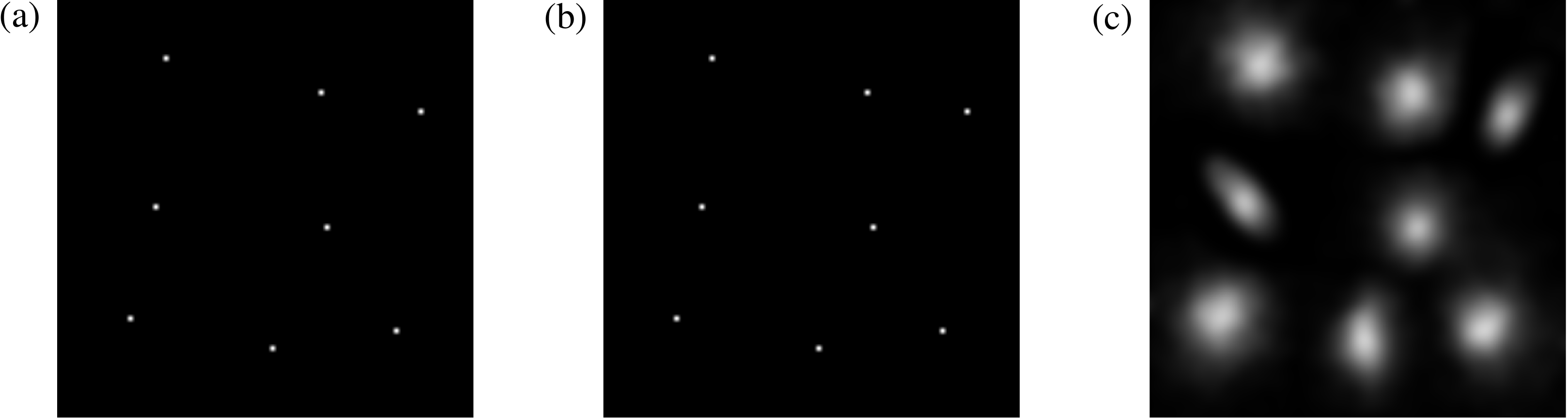}
	\caption{Examples of rendered images of tracers: (a) tracer image rendered in the uniform index field; (b) image of the same tracers for the air flow (C1); (c) image of the same tracers for the water flow (C4). }\label{fig:render}
\end{figure*}

The simulated measurement was carried out in the following order: 
The image of a tracer is rendered at a time $t$, after the ray tracing simulation is performed with the refractive index data which are also from time $t$. The tracer is then moved to a new position following the DNS velocity in a small time interval $\Delta t= 0.005~(L_0/U_0)$, where the two components of the DNS velocity ($u_2$ and $u_3$) are used. The ray tracing simulation of the tracer is performed with the refractive index data from time $t+\Delta t$, after which the image of the same tracer is rendered. 
When the rendered images of a tracer are available at multiple times, the tracer velocity and the acceleration are obtained following the equation~\eqref{eq:velocity} and \eqref{eq:acceleration} with the time interval $\Delta t= 0.005$.

\section{Results and discussion}\label{sec:results}
{In this study, five cases were investigated as summarized in table~\ref{tab:cases}. Because the air and water flow are commonly used in most studies, the results of these two cases are specifically shown and discussed.}
\subsection{Light deflection}\label{sec:theta}
The light deflection $\Delta\pmb \theta={\pmb \theta}_f - {\pmb \theta}_0$ can be written as $\Delta\pmb \theta =  \int [\nabla n/n - (1/n) \cdot (\mathrm{d} n/\mathrm{d} s) \cdot (\mathrm{d} \pmb{r}/\mathrm{d} s)] \mathrm{d} s$, when the integration is taken along the light trajectory with its length $S$ in equation~\eqref{eq:rayeq}. It can be further written as 
\begin{equation}\label{eq:mean_value_eq}
\Delta\pmb \theta = \pmb{N}\cdot S, \quad \mathrm{with} \quad \pmb{N} = \left(\frac{\nabla n}{n} - \frac{1}{n} \frac{\mathrm{d}n}{\mathrm{d}s} \frac{\mathrm{d} \pmb{r}}{\mathrm{d}s} \right)_{\pmb{r}=\xi} ,
\end{equation}
according to \emph{Lagrange mean value theorem} at a point $\xi$ in the ray curve. Here $\pmb{N}$ is determined by the complex ray curve \textit{inside} the flow, and it is difficult (or impossible) to obtain in the non-numerical experiments. The magnitude of the light deflection is $\log(|\Delta\pmb \theta|) = \log (|\pmb N|S) = \log(|\pmb N|L) + \log({S}/{L})$ (according to equation~\ref{eq:mean_value_eq}), where $L$ refers the depth of photon path along the $x_1$ direction.  

Figure~\ref{fig:deltatheta} shows the light deflection obtained in our simulation and that extracted from previous studies, which are summarized in table~\ref{tab:studies}. The light deflection is found to increase linearly with the spatial gradients of the refractive index field, in agreement with previous studies. 
The linear relation can be seen between $|\Delta\pmb \theta|$ and $|\pmb N| L$ (in logarithmic-logarithmic axes) along a gray dash-dot line. 
Our data follow the $|\Delta\pmb \theta|-|\pmb N|L$ trend, and shift as the incidence angle $|\pmb \theta_0|$ increases from $0-0.6$ (red symbols) to $0.6-0.95$ (the blue). Such a shift indicates the link of $|\pmb{\theta}_0|$ to ${S}/{L}$ mentioned above. Figure~\ref{fig:S_L} shows that $S/L$ can be represented by $1/\mathrm{cos} |\pmb{\theta}_0|$, even for the case of the water flow (C5) where the refractive index difference is as large as $10^{-2}$. 

\begin{figure*}
	\centering
	\includegraphics[width=1\textwidth]{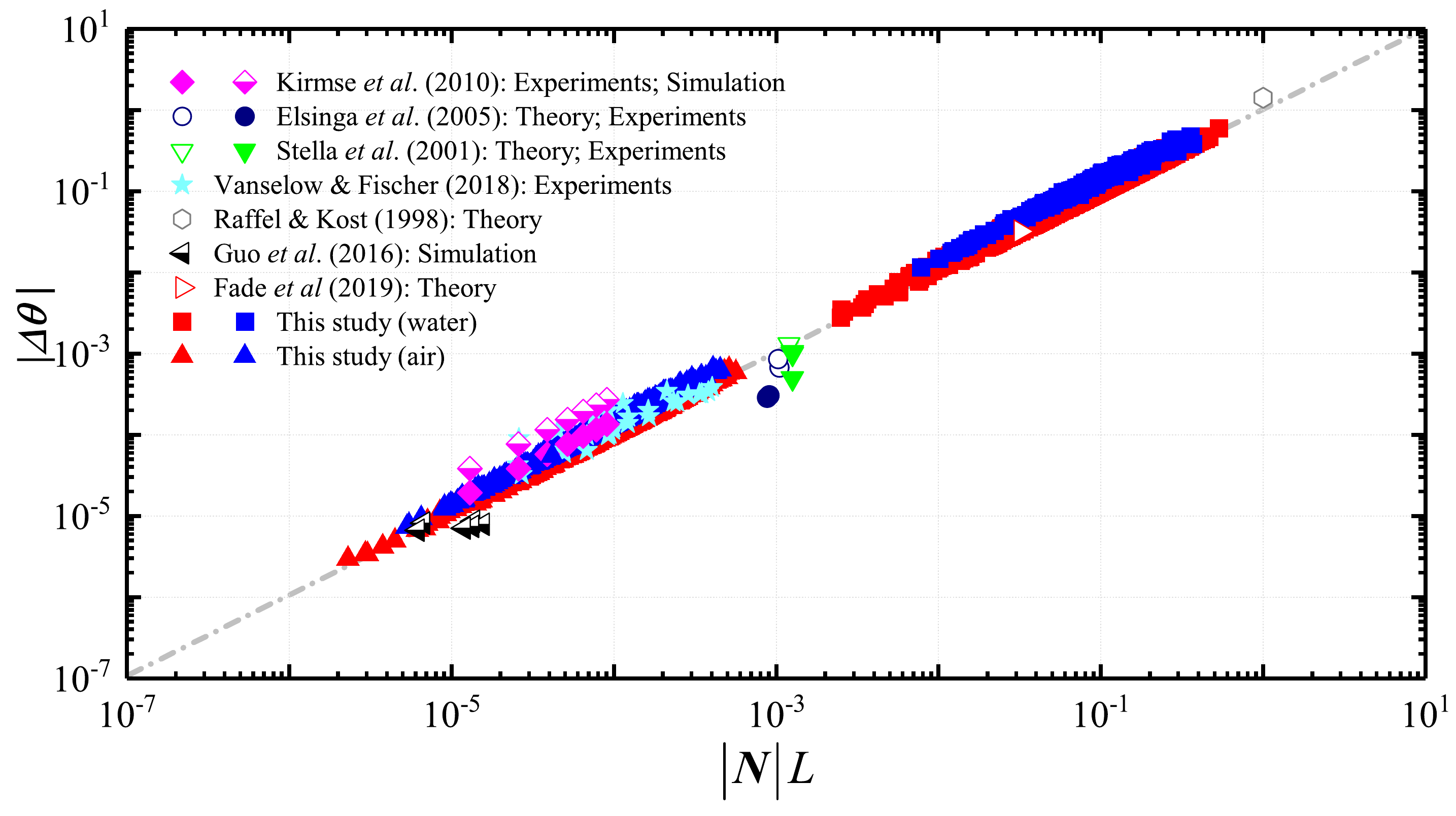}
	\caption{(Color online) Ray deflection $|\Delta \pmb\theta|$ against dimensionless spatial gradients of the refractive indices $|\pmb{N}|L$. For the data of previous studies, $\pmb{N}$ was obtained by $\Delta n_\mathrm{max}$ divided by $n$ and by a characteristic length, and these three values were extracted from each corresponding reference with our best estimation. $\Delta n_\mathrm{max}$ is listed in table~\ref{tab:studies}. The data of $0\leqslant|\pmb\theta_0|\leqslant0.6$ are in blue symbols and those of  $0.6<|\pmb\theta_0|\leqslant0.9$ are in red. Note that $4000$ data points are plotted to avoid oversizing the figure.} \label{fig:deltatheta}
\end{figure*}
\begin{figure*} 
	\centering
	\includegraphics[width=1\textwidth, trim=0cm 0cm 0cm 0cm, clip]{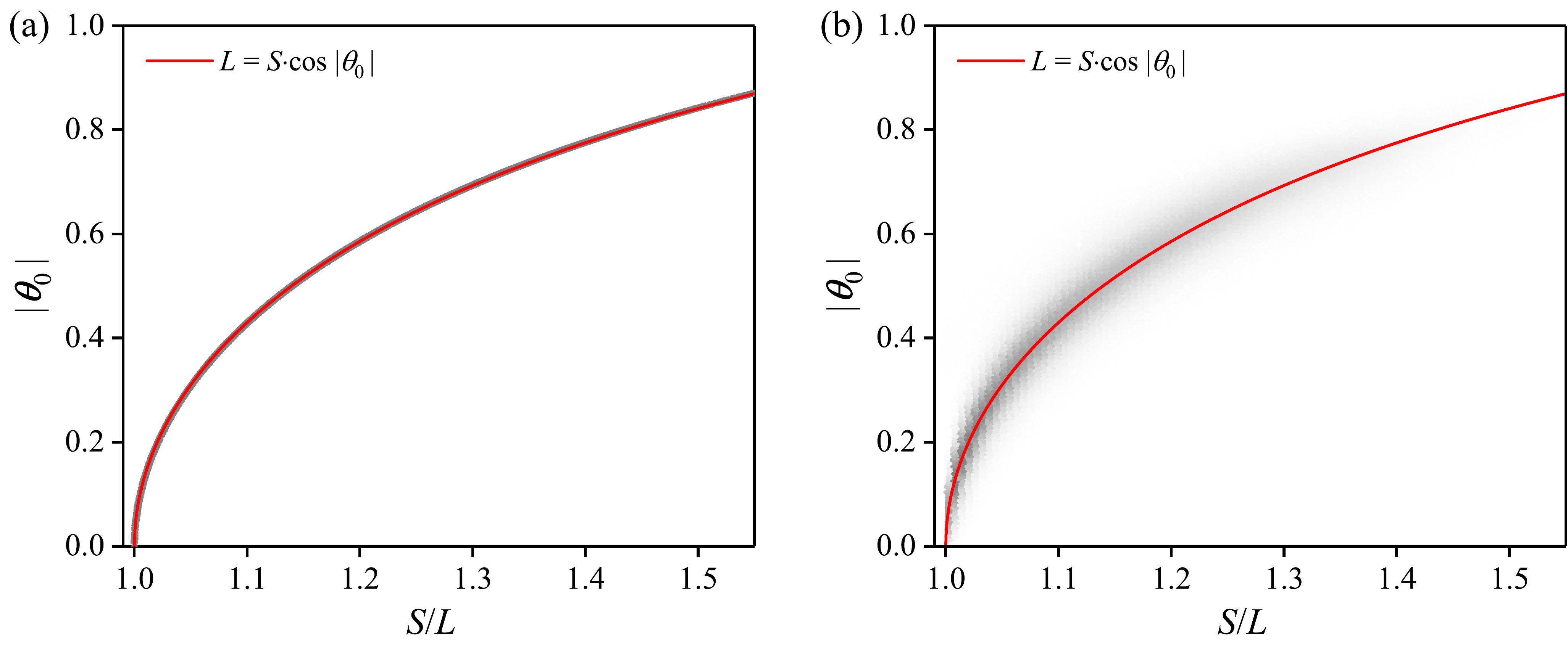}
	\caption{(Color online) The relationship between the length of the light trajectory $S$ and initial incidence angle $|\pmb\theta_0|$: (a) air flow (C1) and (b) water flow (C4). The gray dots denote the sample points, and the grayscale is the two-dimensional probability density function of $|\pmb\theta_0|$ and $S/L$, where darker corresponds to larger value of PDF. } \label{fig:S_L}
\end{figure*}

\subsection{Error of tracer position in images}\label{sec:position}
The effect of the inhomogeneous refractive index field on the imaged tracers has two aspects: one is the position error of the imaged tracer, another one is the shape/area change. The latter is exampled in figure~\ref{fig:render} and detailed statistics of the shape change (quantified by circularity) and the area change are shown in \ref{app:tracer image}.
\begin{figure*}
	\centering
	\includegraphics[width=0.7\textwidth]{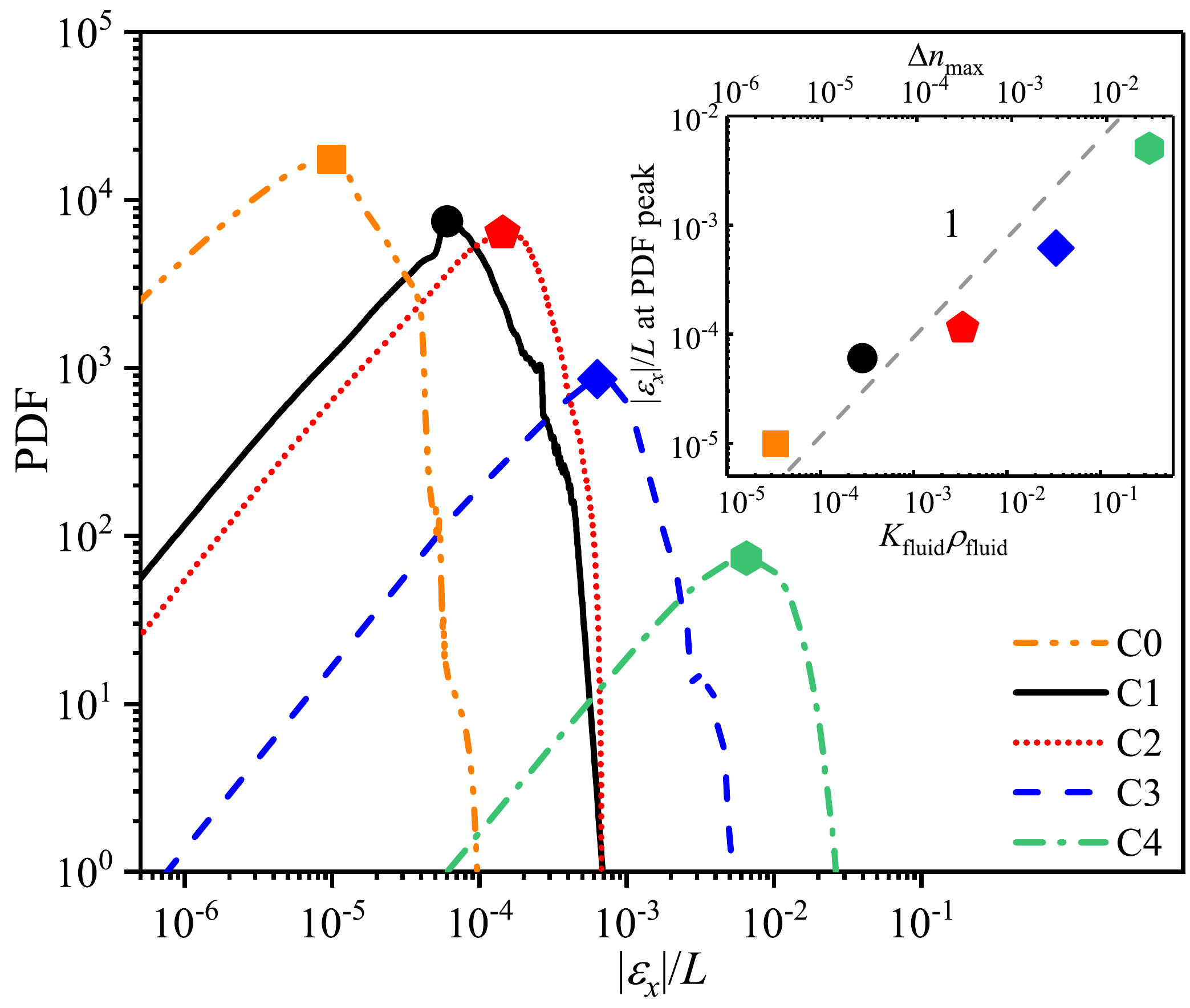}
	\caption{(Color online) PDF distribution of the magnitude of position error $|\pmb{\epsilon}_x|$ for all five cases in this study. The symbols mark the PDF peaks. The inset shows $|\pmb{\epsilon}_x|/L$ at the PDF peaks versus $K_{\mathrm{fluid}} \rho_{\mathrm{fluid}}$ (bottom horizontal axis) and $\Delta n_{\mathrm{max}}$ (top horizontal axis).}\label{fig:shift_pdf}
\end{figure*}

{Regarding to the position error $|\pmb{\epsilon}_x|$, the probability density function (PDF) of $|\pmb{\epsilon}_x|$ is calculated. As shown in figure~\ref{fig:shift_pdf}, {PDF curves of five cases have similar shapes,} and each curve has a peak. The peaks shift towards larger $|\pmb{\epsilon}_x|$ from case C0 to C4. The data at the PDF peaks are extracted and shown in the figure inset, and $|\pmb{\epsilon}_x|$ at the PDF peaks shows an increasing trend with $K_{\mathrm{fluid}} \rho_{\mathrm{fluid}}$. 
Specifically, for the air flow (C1), the magnitudes of position error are up to $\mathcal{O}(10^{-4}L)$, about $1/10$ of the tracer diameter in the images. For the water flow (C4), the magnitudes of position error reach $\mathcal{O}(10^{-2}L)$, corresponding to $\mathcal{O}(10)$ of the diameter of the imaged tracers. 

The ray deflection distance $\pmb{\epsilon}_r = \pmb{r}'-\pmb{r}_0$ ({sketched in figure~\ref{fig:sketch}a}) can be obtained from an integration to the equation~\eqref{eq:mean_value_eq}. $|{\pmb\epsilon}_r|$ can be approximated to be $|\pmb{N} l| \cdot S$ with a length $l$ according to the \emph{Lagrange mean value theorem}, and $|\pmb{N} l|$ could be interpreted as the refractive index mismatch level along the light path. Given that a tracer in the image is resulted from all the rays scattered from a tracer, the position error of the tracer $|{\pmb\epsilon}_x|$ is hence assumed to take the same form as $|{\pmb\epsilon}_r|$,  
\begin{equation}\label{eq:ep_kappa}
|{\pmb\epsilon}_{x}| \approx \widetilde{|\pmb{N}l|} \cdot \widetilde{S/L} \cdot L,
\end{equation}
where $\widetilde{(\cdot)}$ denotes an averaging operation over all $\pmb{r}'$ for one tracer (see figure~\ref{fig:sketch}b). 
\begin{figure*}
	\centering
	\includegraphics[width=1\textwidth, trim=0cm 0cm 0cm 0cm, clip]{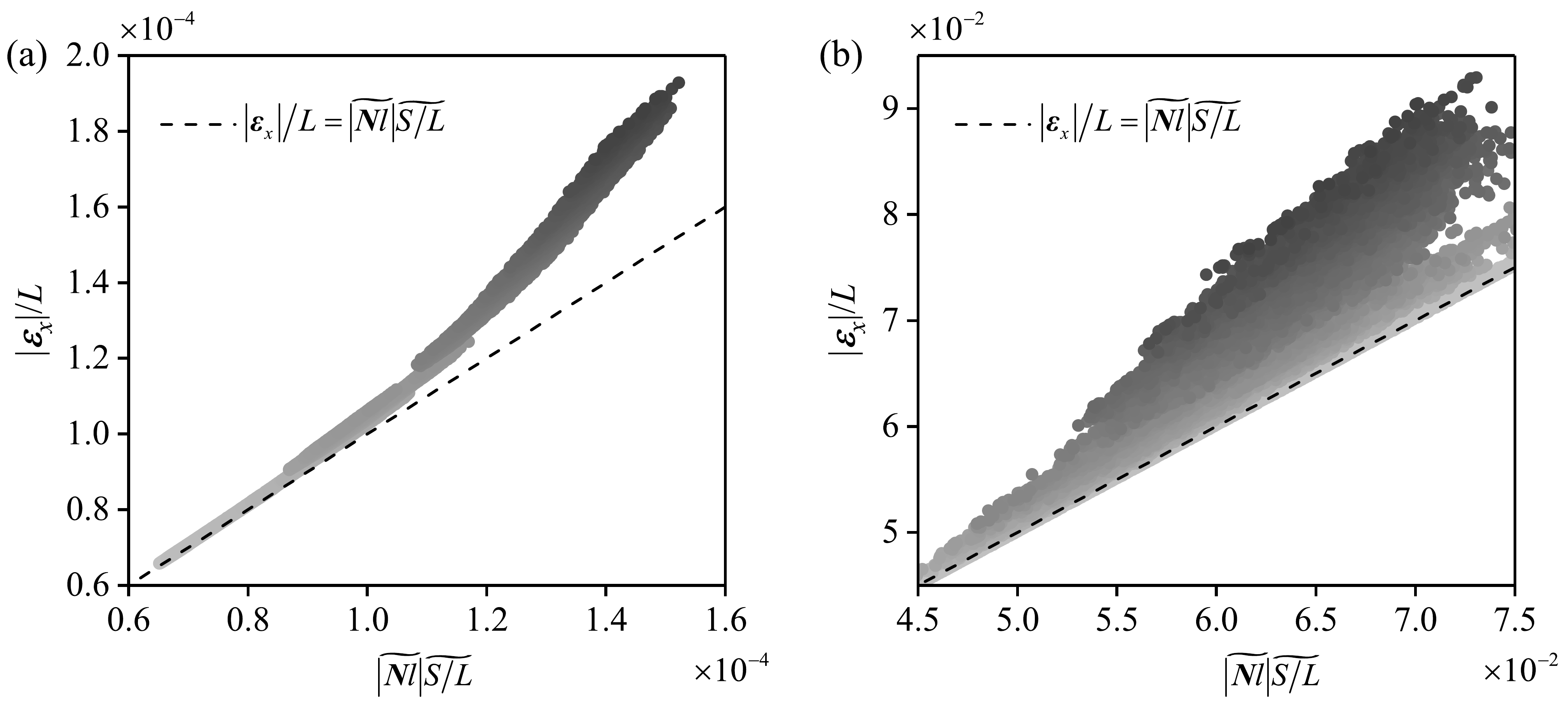}
	\caption{Tracer position error $|{\pmb\epsilon}_x|$ against the relative difference of the refractive indices $\widetilde{|\pmb{N}l|}$: (a) air flow (C1) and (b) water flow (C4). The dots are grayscaled with the incidence angle $\widetilde{\pmb\theta}_0$, the larger $\widetilde{\pmb\theta}_0$ the darker dots.}\label{fig:ep_kappa}
\end{figure*}
{As shown in figure~\ref{fig:ep_kappa}, the tracer position error $|\pmb\epsilon_x|/L$ increases linearly with $\widetilde{|\pmb{N}l|}$ along a dashed line, when $|\pmb\theta_0|$ is close to zero (reaching the paraxial assumption).}
{When $|\pmb\theta_0|$ is increased, the slope of the curve is increased.}
{This suggests that $|\pmb\epsilon_x/L| \approx \widetilde{|\pmb{N}l|}\;\widetilde{S/L}$ $\approx \widetilde{|\pmb{N}l|} /\mathrm{cos}|\widetilde{\pmb\theta}_0|$ (recall that $L \approx S \cdot \mathrm{cos} |\pmb\theta_0|$ shown in figure~\ref{fig:S_L}). }

Based on the above analysis, the position error $|\pmb{\epsilon}_{x}|$ suggests the approximate origination from three aspects: (1) $\widetilde{|\pmb{N}l|}$, which can be referred to the relative difference of the refractive indices (mismatch level) along the light path; (2) $\widetilde{\pmb\theta}_0$, which can be approximately interpreted as the angle of a camera viewing the tracer; (3) $L$, the depth of the index field with which a camera views the tracers.

\subsection{Error of velocity measurement}\label{sec:velocity}

\begin{figure*}
	\centering
	\includegraphics[width=1\textwidth, trim=0 0 0 0, clip]{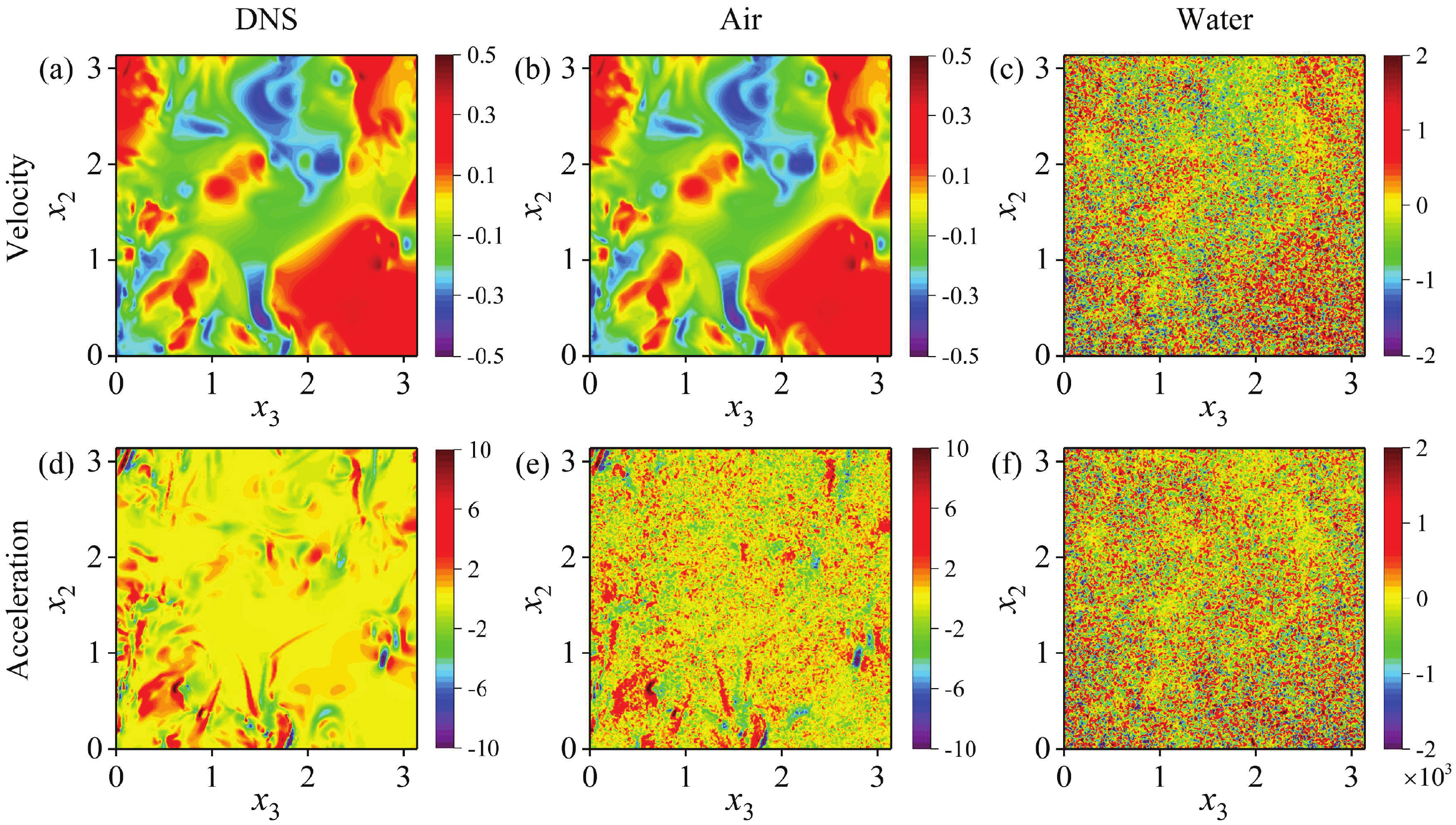}
	\caption{(Color online) Colormap of the flow velocity ($u_3$) and the acceleration ($a_3$) at time $11.4~(L_0/U_0)$ from the DNS (a,d), the air flow case (b, e) and the water flow case (c, f). The velocity and acceleration are made dimensionless with $U_0$ and $U_0^2/L_0$, respectively, as introduced in Section~\ref{sec:dns}.}\label{fig:au_contourf}
\end{figure*}

\begin{figure*}
	\centering
	\includegraphics[width=1\textwidth]{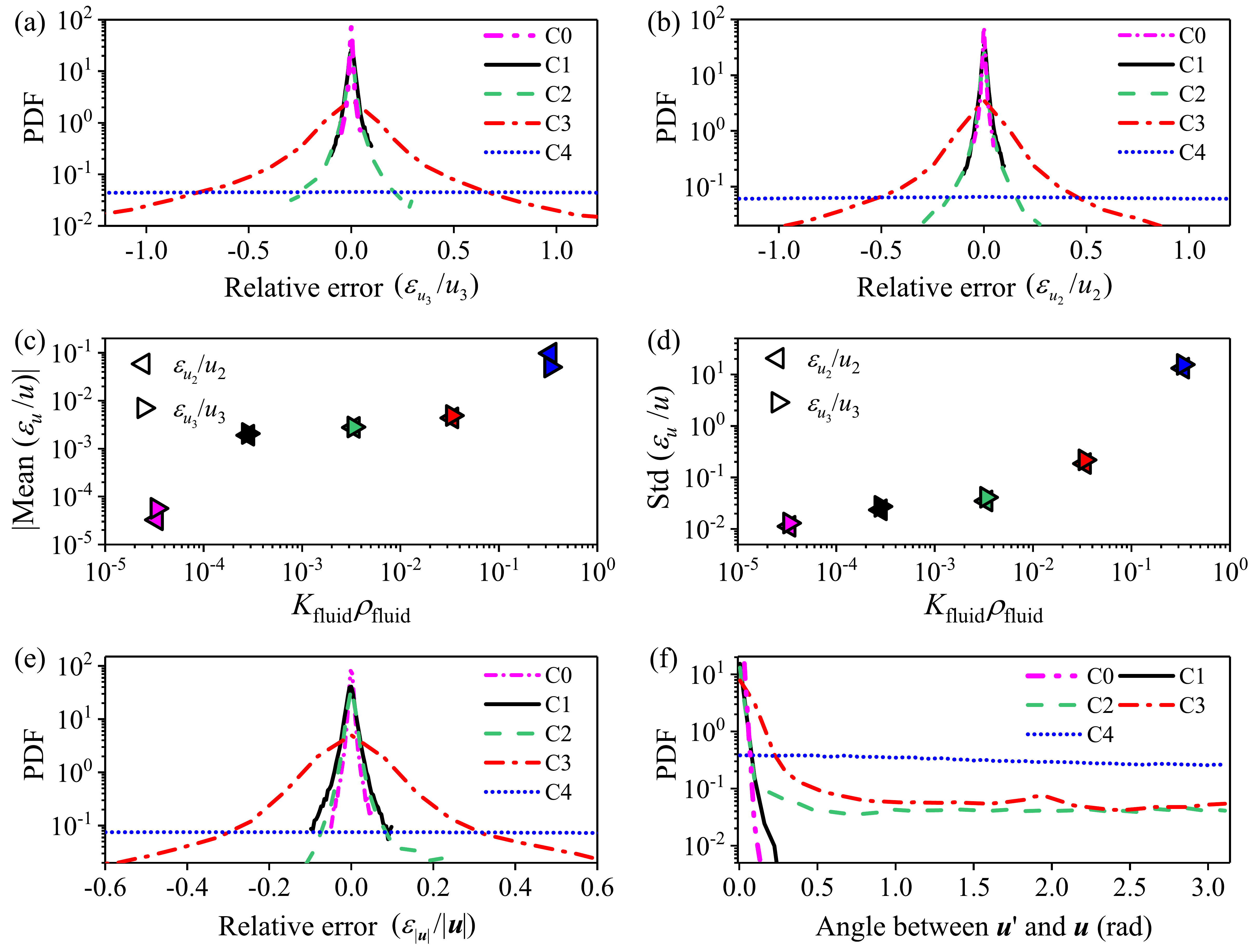}
	\caption{{(Color online) PDF distribution of the relative velocity error $\epsilon_{u_3}/u_{3}$ (a), $\epsilon_{u_2}/u_{2}$ (b), the mean of $\epsilon_{u}/u$ (c), and the corresponding standard deviation (d), the magnitude of the relative error ${\epsilon}_{|\pmb{u}|}/|\pmb{u}|$ (e) and the angle between $\pmb{u}'$ and $\pmb{u}$ (f). In (c) and (d) the filled colors in symbols correspond to the line color of the cases in (a).}}\label{fig:vel}
\end{figure*}

{The contours of the velocity component $u_3$ from the DNS and the simulated measurements for the case C1 and C4 at the same time instant are shown in figure~\ref{fig:au_contourf}(a--c) as examples of visualization. The DNS velocity $\pmb{u}$ is obtained from the cubic interpolation of the DNS data in the $x_2-x_3$ plane, and it is taken as the true for reference. The difference of interpolation schemes on the true value is negligibly small. The velocity contours between the DNS and the air flow (C1) are visually similar, while the velocity contours in the water flow (C4) are fragmented in small scales and contaminated with  large errors. {The velocity component $u_2$ has similar results (not shown).}}

To quantitatively assess the measurement error of the velocity, the PDF statistics is performed for $u_3$ and $u_2$ and their corresponding magnitude and vector direction, as shown in figure~\ref{fig:vel}. The PDF of the relative velocity error $\epsilon_{u_3}/u_{3}$ has a symmetric distribution with its peak close to zero. When $K_{\mathrm{fluid}} \rho_{\mathrm{fluid}}$ is increased, the PDF peak decreases together with broader PDF tails. The PDFs of $\epsilon_{u_2}/u_{2}$ have similar distributions as $\epsilon_{u_3}/u_{3}$. The corresponding mean and standard deviation of $\epsilon_{u_3}/u_{3}$, taken as the \textit{relative systematic measurement error} and \textit{relative random measurement error}, respectively, are shown in panel (c) and (d). The systematic error increases from about $5\times 10^{-3}$ \% to about $10$ \% as $K_{\mathrm{fluid}} \rho_{\mathrm{fluid}}$ is increased. The random error increases from $1$ \% to about $2000$ \% as $K_{\mathrm{fluid}} \rho_{\mathrm{fluid}}$ is increased. The $\epsilon_{u_2}/u_{2}$ has nearly the same systematic and random errors as $\epsilon_{u_3}/u_{3}$. In addition to the statistics of the velocity components, the relative errors of the velocity magnitude is also examined, and the PDFs of $\epsilon_{|\pmb{u}|}/|\pmb{u}|$ have very similar distributions as the components (see panel (e)). 
{The angle between $\pmb{u}'$ and $\pmb{u}$ is obtained by $\mathrm{cos}^{-1}[(\pmb{u}'\cdot\pmb{u})/(|\pmb{u}'||\pmb{u}|)]$, and its PDF has a peak at zero for C0. When $K_{\mathrm{fluid}} \rho_{\mathrm{fluid}}$ is increased, the PDF becomes flatter, especially for the water flow (C4), as shown in figure~\ref{fig:vel}(f).} 

Regarding to the commonly used fluids, air and water, the systematic and the random measurement error of the air flow (C1) is small to be approximately $0.2$~\% and about $2$~\%, respectively, and the flow direction is well measured (see figure~\ref{fig:vel}f). Thus, the effect of the inhomogeneous refractive index might be tolerable (for the turbulent flow in this study). For the water flow (C4), the systematic error reaches $10$ \%, a noticeable level, while the random error reaches about $2000$ \% which demonstrates that the velocity measurement is completely contaminated, so that the measured velocity can be concluded to be questionable, at least for the turbulent flow considered in this study. 
Note that $\Delta n_\mathrm{max} \sim 10^{-5}$ is around the minimum resolution of a portable refractometer, which can be used in refractive index matching techniques to reduce the measurement errors caused by the refractive index difference \citep{xu2012,Dijksman2012,bai2014}. 

The measurement error of the velocity originates from the position error of the tracer due to the light deflection. Studying this error propagation chain, the Lagrangian velocity error is derived (with details in~\ref{app:error_source}) and can be approximately ascribed to $\partial^2 n/\partial t\partial \pmb{x}$ and $\pmb{u}\,\partial^2 n/\partial \pmb{x}^2$, and which the dominant is in variety of flows needs investigations in future. The latter term refers to the advection of the refractive index inhomogeneity, and the former term illustrates that the evolution rate of the refractive index field to the velocity field is a key factor for the velocity measurement error. This evolution rate is often flow dependent. When Taylor's frozen-flow hypothesis is approximately valid, $\partial n/\partial t \approx \pmb{u} \cdot \partial n/\partial \pmb{x}$, and the two error sources turn to be a single one, either $\partial^2 n/\partial t\partial \pmb{x}$ or $\pmb{u}\,\partial^2 n/\partial \pmb{x}^2$. Note that \citet{elsinga2005evaluation} derived the PIV velocity measurement error in an Eulerian scheme, i.e. 
\begin{eqnarray}\label{eq:elsinga}
\pmb{\epsilon}_{u}=\left[\nabla \pmb{\epsilon}_{x}(t)\right] \pmb{u}(t)-[\nabla \pmb{u}(t)] \pmb{\epsilon}_{x}(t),
\end{eqnarray}
where the first part is termed as `the direct velocity error', whereas the second is termed as `the contribution of the position error to the velocity error'. In their equation, the temporal evolution of the refractive index field is not explicitly. However, their expression of the velocity measurement error is indeed consistent with our expression, as shown in detail in \ref{app:error_source}. As a result, the fundamental dependency of the velocity measurement error on the refractive index field is identified and verified.

\subsection{Error of acceleration measurement}\label{sec:acceleration}
\begin{figure*}
	\centering
	\includegraphics[width=1\textwidth]{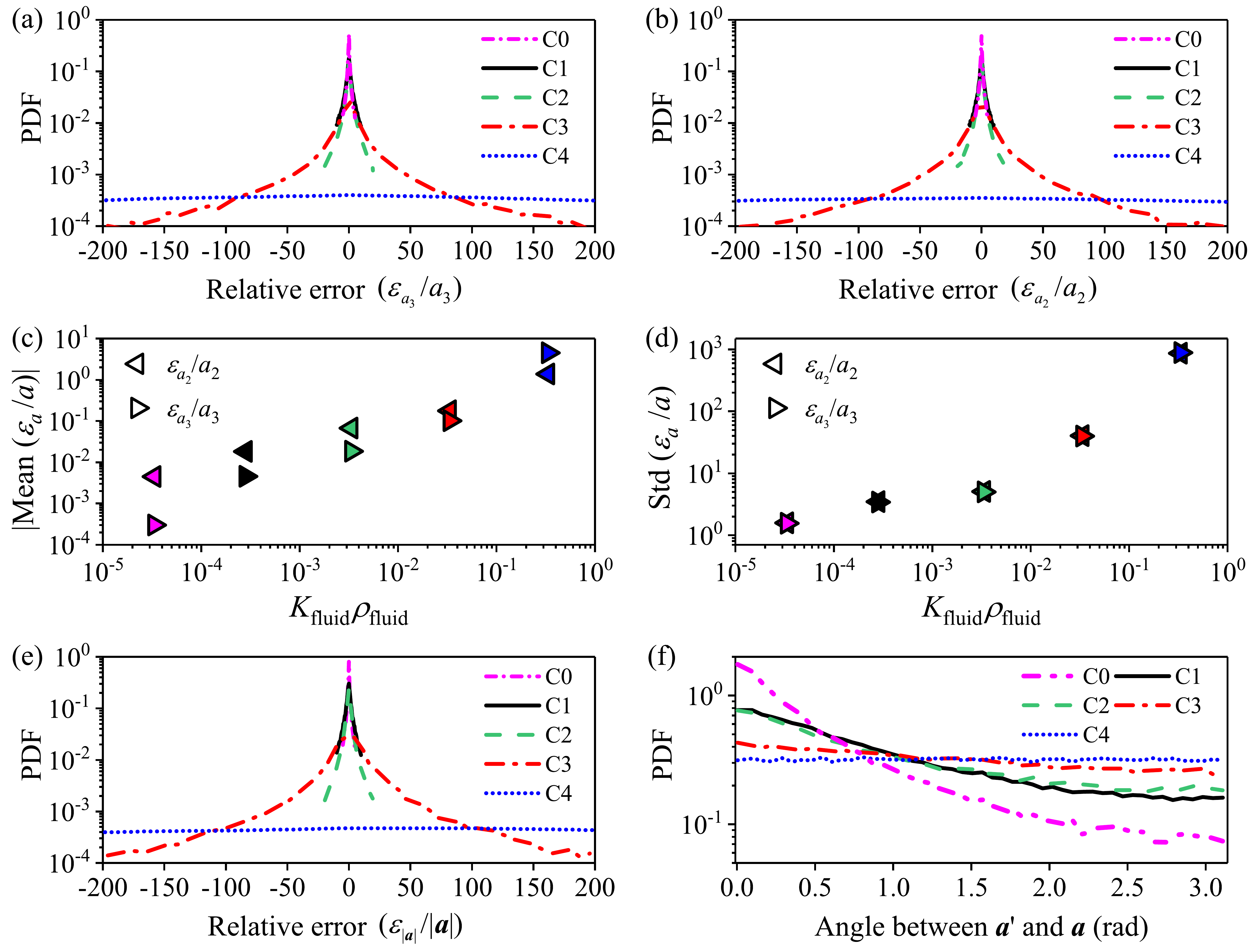}
	\caption{(Color online) PDF distribution of the relative velocity error $\epsilon_{a_3}/a_{3}$ (a), $\epsilon_{a_2}/a_{2}$ (b), the mean of $\epsilon_{a}/a$ (c) and the corresponding standard deviation (d), the magnitude of the relative error ${\epsilon}_{|\pmb{a}|}/|\pmb{a}|$ (e) and the angle between $\pmb{a}'$ and $\pmb{a}$ (f).  In (c) and (d) the filled colors in symbols correspond to the line color of the cases in (a).}\label{fig:acce}
\end{figure*}

The contours of the measured acceleration in the air flow (C1), shown in figure~\ref{fig:au_contourf}(e), are visually similar to the pattern of the DNS data (see figure~\ref{fig:au_contourf}d), although piecemeal `noise' is evidenced. However, for the water case (C4), the contours are significantly contaminated, see figure~\ref{fig:au_contourf}(f).
The PDF distributions of the relative acceleration error ($\epsilon_{a_3}/a_3$) are symmetric with their peaks close to zero, as shown in figure~\ref{fig:acce}(a). When $K_{\mathrm{fluid}} \rho_{\mathrm{fluid}}$ is increased, the distribution peak decreases and the width of the distribution tails becomes broader. The PDFs of $\epsilon_{a_2}/a_{2}$ have similar distributions as $\epsilon_{a_3}/a_{3}$. 
The relative systematic and random errors of the acceleration measurements are evaluated. The systematic error, quantified by the mean of $\epsilon_{a}/a$, increases from about $0.5$ \% to about $100$ \% for $a_2$ and from about $2\times10^{-2}$ \% to about $500$ \% for $a_3$, respectively, as $K_{\mathrm{fluid}} \rho_{\mathrm{fluid}}$ increases. The random error, quantified by standard deviation of $\epsilon_{a}/a$, is found to increase from about $100$ \% to about $10^5$ \% for both $a_2$ and $a_3$ as $K_{\mathrm{fluid}} \rho_{\mathrm{fluid}}$ increases. 
The PDFs of $\epsilon_{|\pmb{a}|}/|\pmb{a}|$ have very similar distributions as the components (see panel (e)). 
The angle between $\pmb{a}'$ and $\pmb{a}$ is shown in panel (f). When $K_{\mathrm{fluid}} \rho_{\mathrm{fluid}}$ is increased, the PDF becomes flatter, as shown in figure~\ref{fig:acce}(f). 

For the case of the air flow (C1), the systematic and the random error of the flow acceleration are about $1$ \% and $300$ \%, respectively. For the case of the water flow (C4), the systematic and the random error are about $500$ \% and $10^5$ \%, respectively.  
Following the same derivation method for the velocity measurement errors, we find that $\pmb{\epsilon}_{a}$ is ascribed to $\partial^2 n/\partial \pmb{x}\partial t$, $\pmb{u}\partial^2 n/\partial \pmb{x}^2$,  $(\partial \pmb{u}/\partial t)(\partial^2 n/\partial \pmb{x}\partial t)$ and $(\partial \pmb{u}/\partial t)(\partial^2 n/\partial \pmb{x}^2)$ (see details in \ref{app:error_source}).

In practice, measurements of flow acceleration are made in three dimensions \citep{Schneiders2016}. In our study, we consider the tracers moving only in the two-dimensional plane ($x_1=0$) for simplicity. Our statistics of two components of the accelerations may give difference to those of three components, but the three-dimensional relative error of acceleration is expected to be on the same level of $\mathcal{O}(1)$ to $\mathcal{O}(10^3)$ for the range of $K_{\mathrm{fluid}} \rho_{\mathrm{fluid}}$ considered here. For the turbulent flow considered in this study, the experimental measurements of the flow acceleration give unrealistic results, even when the maximum refractive index difference is about $10^{-5}$. This finding is from this specific turbulent flow in its flow conditions, but is expected to imply comparable  measurement errors in other flows with similar maximum refractive index difference and similar chaotic levels. 

\section{Conclusion and outlook}\label{sec:conclusion and outlook}
Image quality is crucial for the measurement error of tracer-based velocimetry techniques. When the refractive index field inside a flow is inhomogeneous, tracers in images are blurred and have errors in position, which leads to measurement errors of the flow velocity and the acceleration, respectively. This is particular an issue when the refractive index field is three-dimensional and temporally changes, as in three-dimensional turbulent flows. 

To evaluate the measurement errors in such flows, the distribution of the index field of the flow must be taken into account. 
In this study, ray tracing simulations were carried out to obtain light rays in a single-phase three-dimensional turbulent flow in a simulated experiment. We investigated the flow measurement error regarding every single tracer inside a three-dimensional index field. This field was obtained by converting the DNS density data of the turbulent flow, where two fluids in the same phase mix with each other. Five cases are investigated, with the maximum differences of the refractive indices ranging approximately from $10^{-6}$ to $10^{-2}$. \Reviewsecond{The detailed configuration in the simulation is set for the aim of isolating and focusing on solely evaluating the effect of the inhomogeneous refractive index on the flow measurement errors.}

The measurement errors influenced by the inhomogeneous refractive index field are quantified over four variables: deflection of light rays, position errors of tracers in the image, velocity measurement error and the acceleration measurement error. The analysis of the ray tracing simulation data is in reference to the DNS data of the turbulent flow. The position error of a tracer is found to increase when either the non-dimensional refractive index difference (mismatching level) or the camera viewing angle is increased. This suggests that in preparation of a PTV measurement, the measurement can be refined by reducing the depth of the light rays through the flow field if possible, and/or by reducing the viewing angle of cameras towards the field-of-view.

Regarding to the errors of velocity and acceleration measurements in the turbulent flow considered here, for the case of air flow (with $10^{-5}$ spatial difference of the refractive index field), the relative systematic measurement error is about $0.2$~\% in velocity and about $1$ \% in acceleration, respectively, and this is tolerable (for the turbulent flow in this study). The relative random measurement error is about $2$~\% in velocity and about $300$ \% in acceleration, respectively, i.e., noticeable larger than the systematic error. \Reviewfirst{For the water flow (with $10^{-2}$ spatial refractive index difference), the relative systematic error is about $10$ \% in velocity and about $500$ \% in acceleration, respectively.} The relative random error is about $2000$ \% in velocity and about $10^5$ \% in acceleration, respectively. This clearly shows that, compared with the velocity measurement, the flow acceleration measurement in the water flow is significantly deteriorated to make the measurements untrustworthy (for the turbulent mixing flow in this study). \Reviewfirst{\Fischer{The measurement error is flow dependent, and the measurement error above holds for the studied turbulent mixing flow. Other flows require further investigation, even if their maximum refractive difference is the same, with the proposed methodology.}}
The errors of the velocity and the acceleration are found to be associated with the spatial and spatio-temporal gradients of the refractive index. The latter is controlled by the ratio of velocity-scalar diffusion, whose effects are worthy to be investigated in the future. 

The measurement errors in this work are studied in the framework of PTV. Since PTV and PIV share the same working principle, the findings here are expected to hold for the PIV measurements, which include an averaging effect among the tracer motions within an interrogation window. In addition, the ray tracing simulation method is also applicable for the PTV and PIV techniques with multiple cameras, and other optical flow measurement techniques with working principles based on geometric optics, to evaluate the influence of the inhomogeneous refractive index field on the measurement error.

\section{Acknowledgments}
H. Li gratefully acknowledges the support from Chinese Scholarship Council (No. CSC201804930530). Preliminary work of S. Schuster and T. Schikarski is acknowledged. We thank L. K. Rajendran for answering our questions on the simulation code.

\pagebreak

\appendix
\setcounter{figure}{0}
\section{Verification and convergence tests of the simulation code}\label{app:verification}
{In order to verify {in-house} code, two standard cases in which analytical solutions are available were simulated using our ray tracing script.}
In the first case, a graded-index lens that is often used in optical coupling assemblies was used as the medium. Its two-dimensional refractive index field is
\begin{equation}\label{eq:selfoc_index}
n^2(x_1,x_2) = n_{\mathrm{max}}^2\;(1-0.01^2 x_2^2),
\end{equation}
where $n_{\mathrm{max}}$ is the peak index (see figure~\ref{fig:test}a--b). The analytical solution of a light ray is
\begin{gather}\label{eq:selfoc}
\begin{bmatrix} x_2 \\ \theta \end{bmatrix}
=
\begin{bmatrix} \mathrm{cos}(\alpha x_1) & (1/\alpha)\mathrm{sin}(\alpha x_1) \\ - \alpha\mathrm{sin}(\alpha x_1) & \mathrm{cos}(\alpha x_1) \end{bmatrix}
\begin{bmatrix} x_{2,0} \\ \theta_0 \end{bmatrix}.
\end{gather}
\begin{figure}[t]
	\centering
	\includegraphics[width=1\textwidth]{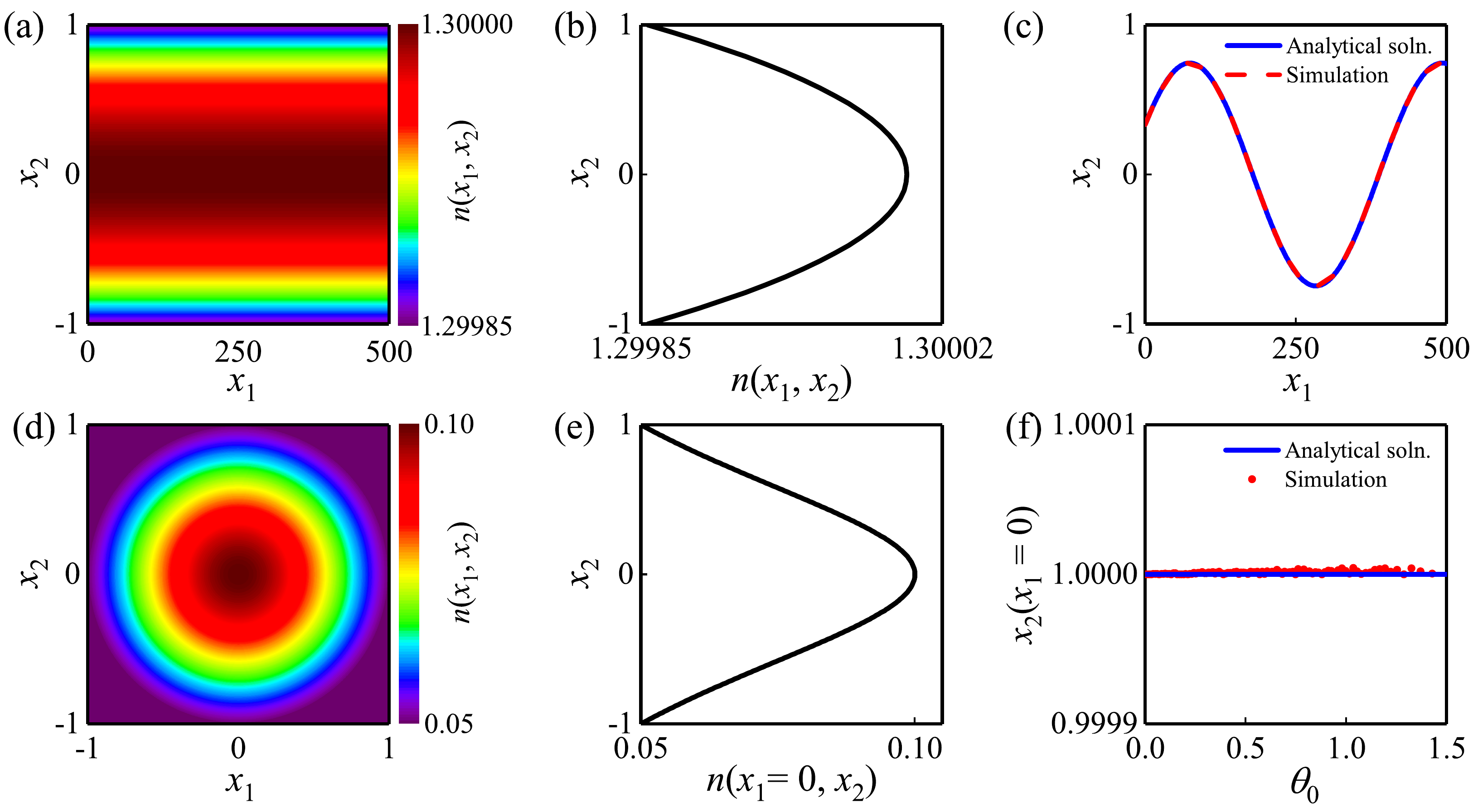}
	\caption{(Color online) Validation of the code: trace rays in a graded-index lens (a)--(c) and a Maxwell's fish-eye lens (d)--(f). (a) Contours of the refractive index field. (b) A profile of the refractive index. (c) The analytical solution and the simulation result. (d) Contours of the refractive index field. (e) The profile of the refractive index at $x_1=0$. (f) The analytical solution and simulation results. Rays are initialized at $(x_1, x_2)=(0,-1)$ with different incident angle $\theta_0$. The feature of the fish-eye lens gives the destination of the rays at $x_{2}|(x_1=0) = 1$ independent on $\theta_0$. }\label{fig:test}
\end{figure}
In figure~\ref{fig:test}(c) the simulated ray agrees well with the analytical solution, and the maximum difference on $x_2$ is up to $5 \times 10^{-5}$. 

In the second case, the light ray in a Maxwell's fish-eye lens, a special example in the family of Luneburg lenses, was simulated. The refractive index of the fish-eye lens is
\begin{equation}\label{eq:Maxwell}
n(x_1,x_2)={n_\mathrm{max}}/[{1+(x_1^2+x_2^2)}],
\end{equation}
as shown in figure~\ref{fig:test}(d)--(e). {The mesh grid of the refractive index field has a number of 512$^2$, whose refractive index also has an approximate range to the water flow case ($\Delta n_{\mathrm{max}} \approx 5 \times 10^{-2}$)}. In a fish-eye lens, a light ray leaving a point on the lens border ends up at a point on the opposite border. The start and the end points are at the same distance from the lens center \citep{maxwell_2011}. Whether this unique characteristic can be reproduced is sensitive to the code precision, thus it is often used for code verification (e.g. in \citep{rajendran2019piv}). In figure~\ref{fig:test}(f), the light rays from $(x_1,x_2)=(0,-1)$ with different incidence angle $\theta_0$ all end up at $(x_1,x_2)=(0,1)$. The maximum difference on $x_2(x_1=0)$ is up to $ 2 \times 10^{-6}$. 


\begin{figure*}
	\centering
	\includegraphics[width=0.9\textwidth]{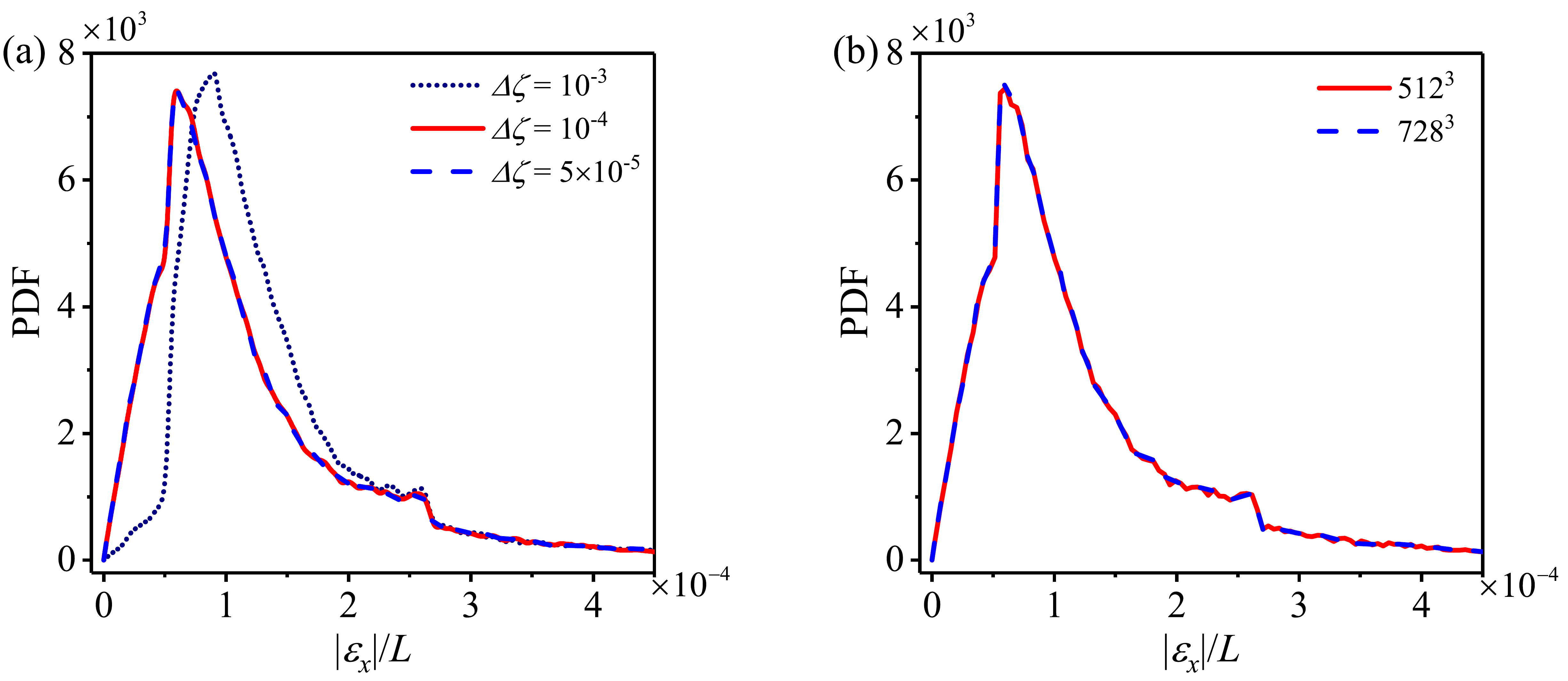}
	\caption{(Color online) PDF distributions of $|\epsilon_x|$ for the case C1 (air) with different simulation step sizes in Runge-Kutta method (a) and with different grid numbers (b).}\label{fig:convergence}
\end{figure*}
{The convergence tests on the simulation time step (in Runge-Kutta iterations) and grid numbers of the refractive index field were carried out. In figure~\ref{fig:convergence}, we showed that the PDF distribution of tracer position error in air flow. It was found that the statistical results with $\Delta \zeta = 10^{-4}$ and $\Delta \zeta = 5\times10^{-5}$ overlap with each other, suggesting that $\Delta \zeta = 10^{-4}$ can give converged simulation. Regarding the grid numbers, the PDF of the results from grid number $512^3$ and that from a finer mesh ($728^3$) of the refractive index field are collapsed, showing that the grid number $512^3$ is sufficient. Thus, the simulation time step $\Delta \zeta = 10^{-4}$ and the mesh number ($512^3$) were employed in our simulations.}

\section{Tracer shape and area}\label{app:tracer image}
\setcounter{figure}{0}
The tracer position is subjected to the deformation of the rendered tracer shape. Here the geometrical properties of the rendered tracer are discussed. The area change of an imaged tracer is $\Delta A = A'-A$ (in $\mathrm{pixel}^2$), where $A'$ denotes the area in an inhomogeneous refractive index field and $A$ the case in a uniform index field. The shape of imaged tracers was quantified by the roundness of an imaged tracer, i.e., circularity $C={P^2}/({4\pi A'})$, where $P$ is perimeter of the tracer. When $C\approx1$, the tracer is a dot in the image, whereas the tracer is elongated if $C \gtrsim 1$ (see figure~\ref{fig:render}). In the air flow (C1), the tracer area change is small with a peak around $0.05$. 
In the water flow (C4), $\Delta A/A$ is significant (with a peak around $38$), and strong blurring of tracers can be seen in images. 
For the shape of imaged tracers, the effect of the refractive index in the air flow (C1) is trivial and imaged tracers are close to dots. 
{However, in the water flow (C4), the tracers are strongly elongated to ellipses, given that $C$ has a distribution with the peak around $2$ and the maximum about $6$. }
\begin{figure*}
	\centering
	\includegraphics[width=1\textwidth]{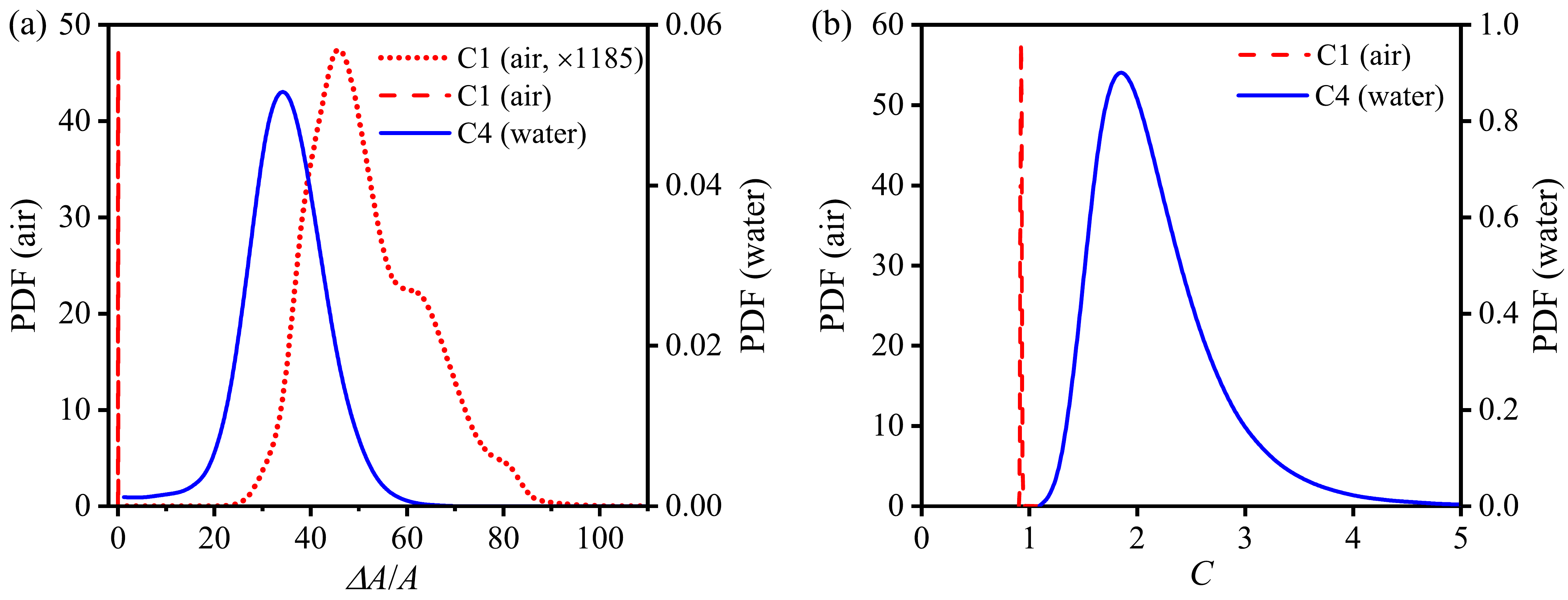}\\
	\caption{(Color online) (a) PDF distribution of change of tracer area $\Delta A$, where $A$ is the tracer area in a uniform refractive index field. The $\Delta A/A$ of air was amplified by $K_\mathrm{water}\rho_{\mathrm{water}}/(K_\mathrm{air}\rho_{\mathrm{air}})\approx1185$ to match the data range of the water for comparison, see the red dotted line. (b) PDF distribution of the circularity $C$ of the imaged tracers.}
	\label{fig:area}
\end{figure*}

\section{Sources of measurement errors}\label{app:error_source}
\setcounter{figure}{0}
In order to locate the sources of the velocity error, evaluating equation~\eqref{eq:velocity_err} with equation~\eqref{eq:velocity} and equation~\eqref{eq:ep_kappa} (as well as approximating $\pmb{\epsilon}_x\approx \mathcal{N} L$, where $\mathcal{N}$ takes the place of  $\widetilde{|\pmb{N}l|} \cdot \widetilde{S/L}$ for simplicity) leads to
\begin{eqnarray}
\label{eq:velocityerror}
\pmb{\epsilon}_u &\approx& [\pmb{x}'(t+\Delta t) - \pmb{x}'(t) - \pmb{x}(t+\Delta t) + \pmb{x}(t)]/\Delta t  \\
&=& [\pmb{x}'(t+\Delta t)-\pmb{x}(t+\Delta t)]/\Delta t - [\pmb{x}'(t)-\pmb{x}(t)]/\Delta t \nonumber \\
&\approx& \{\mathcal{N}[\pmb{x}(t+\Delta t), t+\Delta t] - \mathcal{N}[\pmb{x}(t), t]\} L/\Delta t \nonumber \\
&=& \underbrace{\{\mathcal{N}[\pmb{x}(t+\Delta t), t+\Delta t] - \mathcal{N}[\pmb{x}(t+\Delta t), t]\} L/\Delta t}_\text{\normalsize $ L \cdot\partial \mathcal{N}[\pmb{x}(t+\Delta t)]/\partial t \quad \sim \quad L^2\cdot \partial^2 n/\partial t\partial \pmb{x}$} \nonumber \\
&& +\underbrace{\{\mathcal{N}[\pmb{x}(t+\Delta t), t] - \mathcal{N}[\pmb{x}(t), t]\} L/\Delta t}_\text{\normalsize $ L \cdot \partial \pmb{x}/\partial t \cdot\partial \mathcal{N}(t)/\partial \pmb{x} \quad \sim \quad \pmb{u}L^2\cdot \partial^2 n/\partial \pmb{x}^2 $}\quad . \nonumber
\end{eqnarray}
For the approximation (under the curly bracket), $\mathcal{N}$ (quantifying the refractive index mismatching level, and having the same unit as the refractive index) is approximated by $L\cdot \partial n/\partial \pmb{x}$. 

For detailed consideration of the velocity measurement error derived in \citet{elsinga2005evaluation}, the equation~\eqref{eq:elsinga} is repeated here for convenience,
\begin{equation}
	\pmb{\epsilon}_u=[\nabla \pmb{\epsilon}_x(t)]\pmb{u}(t) - [\nabla \pmb{u}(t)]\pmb{\epsilon}_x(t). \nonumber
\end{equation}
The first term can be approximated as 
\begin{eqnarray}
	\left[\nabla \pmb{\epsilon}_{x}(t)\right] \pmb{u}(t) \approx [L \cdot \partial \mathcal{N}/\partial \pmb{x}] \cdot \pmb{u} \quad \sim \quad \pmb{u} L^2\cdot \partial^2 n/\partial \pmb{x}^2,
\end{eqnarray}
with $\pmb{\epsilon}_x\approx \mathcal{N}L \sim L^2\cdot \partial n/\partial \pmb{x}$ as used above.
The second term can be re-written with taking $\pmb{u}=\partial \pmb{x}/\partial t = (\partial \pmb{x}/\partial n) \cdot (\partial n/\partial t)$, 
\begin{eqnarray}
	\label{eq:elsinga_2term}
	[\nabla \pmb{u}(t)] \cdot \pmb{\epsilon}_{x}(t) &=&\partial \pmb{u}/\partial \pmb{x} \cdot \pmb{\epsilon}_{x}=\partial\left[(\partial n/{\partial t}) \cdot (\partial \pmb{x}/{\partial n})\right]/\partial \pmb{x} \cdot \pmb{\epsilon}_{x} \\ \nonumber
	&\approx& {\partial\left[\partial \pmb{x}/\partial n\right]/\partial \pmb{x} \cdot (\partial n/\partial t) \cdot \mathcal{N} L} +{\partial\left[\partial n/\partial t\right]/\partial \pmb{x} \cdot (\partial \pmb{x}/\partial n) \cdot \mathcal{N} L}. \nonumber
\end{eqnarray}
With $\mathcal{N}\sim L\cdot \partial n/\partial \pmb{x}$, the first term on the right-hand side of equation~\eqref{eq:elsinga_2term} is approximated as $-L^2 \cdot \partial^2 n/ \partial \pmb{x}^2 \cdot (\partial n /\partial t) / (\partial n/\partial \pmb{x})$, which is $-\pmb{u} L^2\cdot \partial^2 n/\partial \pmb{x}^2$, while the second term on the right-hand side of equation~\eqref{eq:elsinga_2term} is approximated as $L^{2} \cdot \partial^{2} n / \partial t \partial \pmb{x}$.

In a short summary, the above approximation shows that the velocity measurement error obtained in  \citet{elsinga2005evaluation}, formulated in equation~\eqref{eq:elsinga}, is also associated with $\partial^2 n/\partial t\partial \pmb{x}$ and $\pmb{u}\,\partial^2 n/\partial \pmb{x}^2$, the same as the finding in our study.

The effect of the velocity and the refractive index fields on the measurement error of flow acceleration can be approximated as follows,
\begin{eqnarray}
\pmb{\epsilon}_{a} &=& \pmb{a}' - \pmb{a}  \\
&=& {\partial \pmb{u}'}/{\partial t} + (\pmb{u}' \cdot \nabla) \pmb{u}' - {\partial \pmb{u}}/{\partial t} - (\pmb{u} \cdot \nabla) \pmb{u} \nonumber \\
&=& \underbrace{{\partial \pmb{u}'}/{\partial t} - {\partial \pmb{u}}/{\partial t}}_\text{\normalsize$\pmb{\epsilon}_{a,I}$} + \underbrace{(\pmb{u}' \cdot \nabla) \pmb{u}' - (\pmb{u} \cdot \nabla) \pmb{u}}_\text{\normalsize$\pmb{\epsilon}_{a,II}$} \nonumber 
\end{eqnarray}
The two components of $\pmb{\epsilon}_{a}$ are approximated individually as follows,
\begin{flalign}
\pmb{\epsilon}_{a,I} \approx& (1/\Delta t)^2[\pmb{x}'(t+2\Delta t) - 2\pmb{x}'(t+\Delta t) + \pmb{x}'(t)] \\
&- (1/\Delta t)^2[\pmb{x}(t+2\Delta t) - 2\pmb{x}(t+\Delta t)  + \pmb{x}(t)]\nonumber\\
\approx& (1/\Delta t)^2[\pmb{x}'(t+2\Delta t) - \pmb{x}(t+2\Delta t) - 2\pmb{x}'(t+\Delta t) + 2\pmb{x}(t+\Delta t) + \pmb{x}'(t) - \pmb{x}(t)] \nonumber \\
\approx& \underbrace{(1/\Delta t)\{\mathcal{N}[\pmb{x}(t+2\Delta t), t+2\Delta t] - \mathcal{N}[\pmb{x}(t+2\Delta t), t+\Delta t]\}\cdot L}_\text{\normalsize$ L\cdot\partial \mathcal{N}[\pmb{x}(t+2\Delta t)]/\partial t \; \sim \; L^2\cdot\partial^2 n/\partial t\partial \pmb{x}$} /\Delta t \nonumber\\
& + \underbrace{(1/\Delta t)\{\mathcal{N}[\pmb{x}(t+2\Delta t), t+\Delta t] - \mathcal{N}[\pmb{x}(t+\Delta t), t+\Delta t]\}\cdot L}_\text{\normalsize$ \pmb{u}(t+\Delta t)L\cdot\partial \mathcal{N}(t+\Delta t)/\partial \pmb{x} \; \sim \; \pmb{u}L^2\cdot \partial^2 n/\partial \pmb{x}^2$} /\Delta t \nonumber\\
& - \underbrace{(1/\Delta t)\{\mathcal{N}[\pmb{x}(t+\Delta t), t+\Delta t] - \mathcal{N}[\pmb{x}(t+\Delta t), t]\}\cdot L}_\text{\normalsize$ L\cdot\partial \mathcal{N}[\pmb{x}(t+\Delta t)]/\partial t \; \sim \; L^2\cdot\partial^2 n/\partial t\partial \pmb{x}$} /\Delta t \nonumber\\
& - \underbrace{(1/\Delta t)\{\mathcal{N}[\pmb{x}(t+\Delta t), t] - \mathcal{N}[\pmb{x}(t), t]\}\cdot L}_\text{\normalsize$ \pmb{u}(t)L\cdot\partial \mathcal{N}[\pmb{x}(t+\Delta t)]/\partial \pmb{x} \; \sim \; \pmb{u}L^2\cdot\partial^2 n/\partial \pmb{x}^2$} /\Delta t \nonumber\\
& \nonumber\\
\sim& \quad \partial^2 n/\partial t\partial \pmb{x} \quad \&\quad \pmb{u} \partial^2 n/\partial \pmb{x}^2  \nonumber
\end{flalign}

\begin{flalign}
\pmb{\epsilon}_{a,II} \approx& (1/\Delta t)^2(1/\Delta \pmb{x}) [\pmb{x}'(t+\Delta t) - \pmb{x}'(t)][\pmb{x}'(t+2\Delta t) - 2\pmb{x}'(t+\Delta t) + \pmb{x}'(t)] \\
&-(1/\Delta t)^2(1/\Delta \pmb{x})[\pmb{x}(t+\Delta t) - \pmb{x}(t)][\pmb{x}(t+2\Delta t) - 2\pmb{x}(t+\Delta t) + \pmb{x}(t)] \nonumber\\
\approx& (1/\Delta t)^2(1/\Delta \pmb{x}) [\pmb{x}'(t+\Delta t) - \pmb{x}'(t)][\pmb{x}'(t+2\Delta t) - 2\pmb{x}'(t+\Delta t) + \pmb{x}'(t)] \nonumber \\
& -(1/\Delta t)^2(1/\Delta \pmb{x}) [\pmb{x}'(t+\Delta t) - \pmb{x}'(t)][\pmb{x}(t+2\Delta t) - 2\pmb{x}(t+\Delta t) + \pmb{x}(t)] \nonumber \\
& +(1/\Delta t)^2(1/\Delta \pmb{x}) [\pmb{x}'(t+\Delta t) - \pmb{x}'(t)][\pmb{x}(t+2\Delta t) - 2\pmb{x}(t+\Delta t) + \pmb{x}(t)] \nonumber \\
&-(1/\Delta t)^2(1/\Delta \pmb{x})[\pmb{x}(t+\Delta t) - \pmb{x}(t)][\pmb{x}(t+2\Delta t) - 2\pmb{x}(t+\Delta t) + \pmb{x}(t)] \nonumber\\
\approx& (1/\Delta t)^2(1/\Delta \pmb{x}) [\pmb{x}'(t+\Delta t) - \pmb{x}'(t)][\pmb{x}'(t+2\Delta t) - \pmb{x}(t+2\Delta t)\nonumber \\ 
&- 2\pmb{x}'(t+\Delta t) + 2\pmb{x}(t+\Delta t)+ \pmb{x}'(t)-\pmb{x}(t)] + (1/\Delta t)^2(1/\Delta \pmb{x}) [\pmb{x}'(t+\Delta t) \nonumber\\
&-\pmb{x}(t+\Delta t) - \pmb{x}'(t) + \pmb{x}(t)][\pmb{x}(t+2\Delta t) - 2\pmb{x}(t+\Delta t) + \pmb{x}(t)]\nonumber \\
\approx& (1/\Delta t)(1/\Delta \pmb{x}) (L) \pmb{u}'(t)\{\mathcal{N}[\pmb{x}(t+2\Delta t), t+2\Delta t] - \mathcal{N}[\pmb{x}(t+\Delta t), t+\Delta t] \nonumber \\
&+ \mathcal{N}[\pmb{x}(t), t]\} + (1/\Delta t)^2(1/\Delta \pmb{x})(L) \{\mathcal{N}[\pmb{x}(t+\Delta t), t+\Delta t] -\mathcal{N}[\pmb{x}(t), t] \}\nonumber \\
& \cdot[\pmb{x}(t+2\Delta t)- 2\pmb{x}(t+\Delta t) + \pmb{x}(t)]\nonumber 
\end{flalign}
\begin{flalign}
\approx & \; \pmb{u}'(t)\left[\underbrace{\{\mathcal{N}[\pmb{x}(t+2\Delta t), t+2\Delta t] - \mathcal{N}[\pmb{x}(t+2\Delta t), t+\Delta t]\}L/\Delta t}_\text{\normalsize$ L\cdot\partial \mathcal{N}[\pmb{x}(t+2\Delta t)]/\partial t \; \sim \; L^2\cdot\partial^2 n/\partial t\partial \pmb{x}$}(1/\Delta \pmb{x}) \right.  \nonumber\\
& + \{\underbrace{\mathcal{N}[\pmb{x}(t+2\Delta t), t+\Delta t]  - \mathcal{N}[\pmb{x}(t+\Delta t), t+\Delta t]\}(L/\Delta \pmb{x}) }_\text{\normalsize$ L\cdot\partial \mathcal{N}(t+\Delta t)/\partial \pmb{x} \; \sim \; L^2\cdot\partial^2 n/\partial \pmb{x}^2$}(1/\Delta t)  \nonumber \\
& - \underbrace{\{\mathcal{N}[\pmb{x}(t+\Delta t), t+\Delta t] - \mathcal{N}[\pmb{x}(t+\Delta t), t]\}(L/\Delta t)}_\text{\normalsize$ L\cdot\partial \mathcal{N}[\pmb{x}(t+\Delta t)]/\partial t \; \sim \; L^2\cdot\partial^2 n/\partial t\partial \pmb{x}$}(1/\Delta \pmb{x}) \nonumber\\
&\left. - \underbrace{\{\mathcal{N}[\pmb{x}(t+\Delta t), t] - \mathcal{N}[\pmb{x}(t), t]\}(L/\Delta \pmb{x}) }_\text{\normalsize$ L\cdot\partial \mathcal{N}(t)/\partial \pmb{x} \; \sim \; L^2\cdot\partial^2 n/\partial \pmb{x}^2$}(1/\Delta t)  \right]\nonumber \\
& + \frac{\partial \pmb{u}}{\partial t} (\Delta t)\left[\underbrace{\{\mathcal{N}[\pmb{x}(t+\Delta t), t+\Delta t] - \mathcal{N}[\pmb{x}(t+\Delta t), t]\}(L/\Delta t)}_\text{\normalsize$ L\cdot \partial \mathcal{N}[\pmb{x}(t+\Delta t)]/\partial t \; \sim \; L^2\cdot \partial^2 n/\partial \pmb{x}\partial t$}(1/\Delta \pmb{x}) \right.\nonumber \\
&\left.+ \underbrace{\{\mathcal{N}[\pmb{x}(t+\Delta t), t]-\mathcal{N}[\pmb{x}(t), t]\}(L/\Delta \pmb{x}) }_\text{\normalsize$ L\cdot \partial \mathcal{N}(t)/\partial \pmb{x} \; \sim \; L^2\cdot\partial^2 n/\partial \pmb{x}^2$}(1/\Delta t) \right]\nonumber\\ 
& \nonumber\\
\sim & \quad \pmb{u}'\partial^2 n/\partial t\partial \pmb{x} \quad \& \quad (\partial \pmb{u}/\partial t)(\partial^2 n/\partial t\partial \pmb{x} + \partial^2 n/\partial \pmb{x}^2) 
\end{flalign}
In summary, the measurement error of the flow acceleration $\pmb{\epsilon}_{a}$ is approximately ascribed to $\partial^2 n/\partial \pmb{x}\partial t$, $\pmb{u}\partial^2 n/\partial \pmb{x}^2$,  $(\partial \pmb{u}/\partial t)(\partial^2 n/\partial \pmb{x}\partial t)$ and $(\partial \pmb{u}/\partial t)(\partial^2 n/\partial \pmb{x}^2)$, spatial and spatio-temporal gradients of refractive index field.

\bibliographystyle{elsarticle-num-names} 
\bibliography{raytracing}


%
%
%
\end{document}